\documentclass[%
 reprint,
 amsmath,amssymb,
 aps,
 longbibliography,
]{revtex4-1}

\usepackage{graphicx}
\usepackage{dcolumn}
\usepackage{bm}

\begin{document}

\preprint{APS/123-QED}

\title{Single-shot x-ray speckle-based imaging of a single-material object}

\author{Konstantin M. Pavlov}
\email{konstantin.pavlov@canterbury.ac.nz}
\affiliation{School of Physical and Chemical Sciences, University of Canterbury, Christchurch, New Zealand}
\affiliation{School of Physics and Astronomy, Monash University, Victoria 3800, Australia}
\affiliation{School of Science and Technology, University of New England, NSW 2351, Australia}

\author{Heyang (Thomas) Li}
\affiliation{School of Mathematics and Statistics, University of Canterbury, Christchurch, New Zealand}
\affiliation{School of Physics and Astronomy, Monash University, Victoria 3800, Australia}

\author{David M. Paganin}
\affiliation{School of Physics and Astronomy, Monash University, Victoria 3800, Australia}

\author{Sebastien Berujon}
\affiliation{European Synchrotron Radiation Facility, 38043 Grenoble,
France}

\author{H\'{e}l\`{e}ne Roug\'{e}-Labriet}
\affiliation{Novitom, 3 av doyen Louis Weil 38000 Grenoble, France}
\affiliation{Inserm UA7 STROBE, Universit\'{e} Grenoble Alpes, 38000 Grenoble, France}

\author{Emmanuel Brun}
\affiliation{Inserm UA7 STROBE, Universit\'{e} Grenoble Alpes, 38000 Grenoble, France}

\date{\today}

\begin{abstract}
We develop a means for speckle-based phase imaging of the projected thickness of a single-material object, under the assumption of illumination by spatially random time-independent x-ray speckles. These speckles are generated by passing x rays through a suitable spatially random mask. The method makes use of a single image obtained in the presence of the object, which serves to deform the illuminating speckle field relative to a reference speckle field (which only needs to be measured once) obtained in the presence of the mask and the absence of the object. The method implicitly rather than explicitly tracks speckles, and utilizes the transport-of-intensity equation to give a closed-form solution to the inverse problem of determining the complex transmission function of the object. Implementation using x-ray synchrotron data shows the method to be robust and efficient with respect to noise. Applications include x-ray phase--amplitude radiography and tomography, as well as time-dependent imaging of dynamic and radiation-sensitive samples using low-flux sources.
\end{abstract}

\maketitle


\section{\label{sec:intro}Introduction}

X-ray Phase Contrast Imaging (PCI) (see e.g.~\citeauthor{paganin2006} \cite{paganin2006} and references therein) appeared a few decades ago and quickly demonstrated higher diagnostic potential compared to conventional attenuation-based tomography. With the emergence of partially coherent x-ray sources over twenty years ago, expectations regarding PCI became achievable with several methods developed mostly at synchrotrons, some of which were later adapted to laboratory sources \cite{bravin2012}. Among all PCI techniques that utilize additional optical elements, Speckle Based Imaging \cite{berujon2012,morgan2012,zdora2018} makes use of a very simple experimental set-up, and has already been demonstrated to be a good candidate for transfer outside synchrotron facilities. The set-up, in addition to its simplicity of implementation, presents the main advantages of having no field of view limitation other than that imposed by the detector (speckle-generating diffusers of any size are easy to manufacture), and relatively low requirements on the beam coherence. As the use of speckle as a wave-front modulator for phase contrast x-ray imaging was proven valuable less than a decade ago \cite{berujon2012,morgan2012}, only a few publications have dealt with the corresponding phase-retrieval problem. 

To detect the refraction of x rays due to their passage through a sample, the speckle technique has been primarily based on tracking transverse speckle displacement \cite{berujon2012,morgan2012,zdora2018} between (i) reference speckle images (i.e.~with a membrane generating a random spatial modulation), and (ii) sample speckle images (i.e.~with the membrane and the sample). Recently a technique was developed for speckle-tracking phase retrieval based on a geometric-flow formalism  \cite{PaganinLabrietBrunBerujon2018}; this approach was  inspired by both the transport-of-intensity equation of paraxial wave optics \cite{teague1983} and the optical-flow formulation of image processing \cite{OF_old1,OF_old2}.  The geometric-flow speckle-tracking method \cite{PaganinLabrietBrunBerujon2018} has demonstrated good efficiency for retrieving phase shifts of a non-absorbing sample using only a single sample exposure. The method implicitly rather than explicitly tracks speckles, and therefore has potentially an added advantage of finer spatial resolution compared to methods that explicitly track speckles using e.g.~correlation windows of at least a few pixels in width and height. The reason for this possibly finer spatial resolution is that, with implicit speckle tracking, the image does not need to be divided into correlation windows.  The geometric-flow formalism works by solving a differential equation (continuity equation) that enforces local energy conservation at every pixel \cite{PaganinLabrietBrunBerujon2018}, and so does not need to define multi-pixel regions over which correlation coefficients are computed.  This geometric-flow method has also very recently been translated to visible-light microscopy \cite{Lu2019}. Here we extend the geometric-flow results of  \citet{PaganinLabrietBrunBerujon2018} to the case of an absorbing monomorphous \cite{Paganin2004,GureyevMonomorphous2015} sample, which by assumption obeys at least one of the following approximations: (i) a single-material sample of possibly variable density \cite{Paganin_single_material_2002}; (ii) a sample for which the logarithm of attenuation is proportional to the corresponding phase shift \cite{Paganin2004,Wu2005}, or equivalently a sample of possibly variable density for which the ratio of the real refractive index decrement, and the imaginary part of the complex refractive index, is constant within the volume occupied by the sample \cite{Paganin2004,GureyevMonomorphous2015};  (iii) a sample composed of light elements ($Z<10$) when irradiated with high-energy x rays (60$-$500 keV) \cite{Wu2005}, this being a special case of the single-material assumption  \cite{Paganin_single_material_2002,Paganin2004} in which the single material may be well approximated as electron density.  As shall be shown below, an x-ray speckle-tracking experiment, that satisfies any one or more of the above assumptions, will be amenable to analysis using the same reconstruction algorithm.

We close this introduction with a brief overview of the remainder of the paper. Section ~\ref{sec:theory} is dedicated to the theoretical development of our method for single-shot x-ray speckle-based imaging of a single-material object. Section \ref{sec:experiment1} describes the x-ray synchrotron experiment, used to obtain data to implement our method for x-ray phase imaging in both two and three spatial dimensions. The results of applying our theory to these data are given in 
Sec.~\ref{sec:experiment2}. We discuss some broader implications of our work, together with possible future extensions and applications, in Sec.~\ref{sec:discussion}. We conclude with a summary in Sec.~\ref{sec:conclusion}.

\section{\label{sec:theory}Theory}

\begin{figure}
\includegraphics[trim=0 0 0 0,clip,width=0.48\textwidth,scale=0.2]{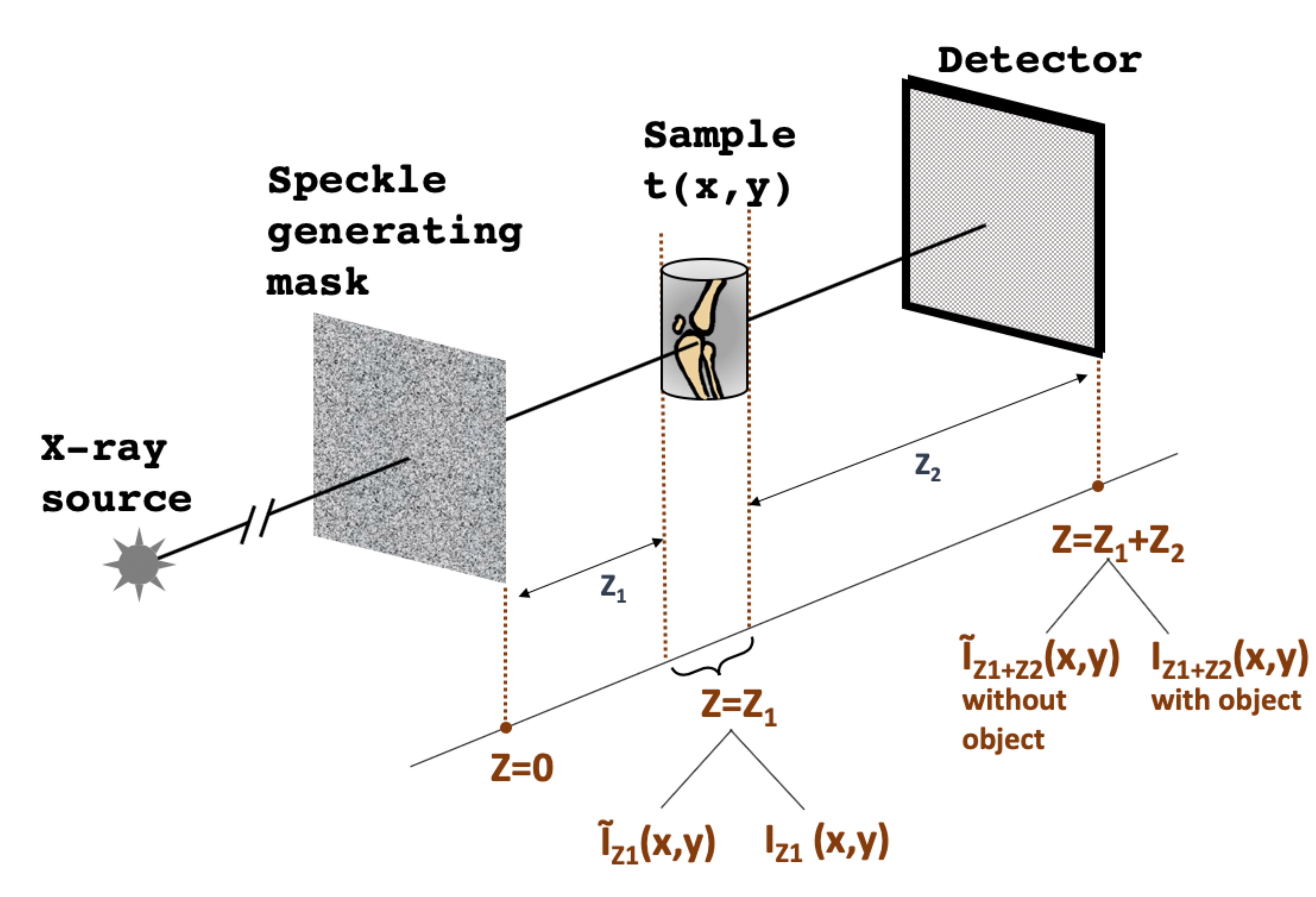}
\caption{Generic setup for x-ray speckle tracking.}
\label{fig:ExperimentalSetUp}
\end{figure}

Consider unit-intensity monochromatic x-ray scalar plane waves, incident on a spatially random mask, which produce a speckle image downstream---see Fig.~\ref{fig:ExperimentalSetUp}. Paraxial radiation is assumed throughout, and we use the term {\em speckle} to refer to any spatially-random intensity distribution, not necessarily corresponding to fully developed speckle \footnote{As shall become clear later in the paper, the spatially-random intensity distributions that we consider are not fully-developed speckle. Much literature equates {\em speckle} with {\em fully developed speckle}, but we adopt a more broadly applicable usage of the term {\em speckle} to refer to any spatially-random intensity distribution.}. If we further assume that the incident plane wave propagates along the positive $z$-axis, which we identify with the optical axis, then such a wave can be represented as $\exp(ikz)$, where $k=2\pi/\lambda$ and $\lambda$ is the wavelength in vacuum. The distance between the mask and the object is $z_1$, and the distance between the object and the downstream position-sensitive detector is $z_2$. We assume the object to have negligible extension in the $z$-direction.

We assume that the transport-of-intensity equation (TIE) \cite{teague1983} approximation is valid for this experiment (see Fig.~\ref{fig:ExperimentalSetUp}). This is because the following validity conditions are, by assumption, met: (i) paraxial x-ray radiation; (ii) thin paraxially-scattering sample; (iii) object-to-detector distance that is sufficiently small for the longitudinal evolution of intensity to be linear with respect to the propagation distance; (iv) negligible inelastic scattering by the sample. Then the longitudinal derivative of the 2D intensity distribution just after the object, $I_{z_1}(x,y)$, with respect to the $z$-coordinate can be estimated via \cite{teague1983}:
\begin{eqnarray}
\label{eq:1}
\frac{\partial I_{z_1}(x,y)}{\partial z} = -\frac{1}{k} \nabla_{\perp}\cdot[I_{z_1}(x,y)\nabla_{\perp}\varphi(x,y)],
\end{eqnarray}
where $\varphi(x,y)$ is the phase profile of the wave-field just after the object, $(x,y)$ are Cartesian coordinates in planes perpendicular to $z$ and $\nabla_{\perp}=(\partial/\partial x,\partial/\partial y)$. Here, the phase shift $\varphi(x,y)$ is caused by both the object (as indicated by the ``ob'' subscript) and the mask (subscript ``m''):
\begin{eqnarray}
\label{eq:7}
\varphi(x,y)=\varphi_{\textrm{ob}}(x,y)+\varphi_{\textrm{m}}(x,y).
\end{eqnarray}

For large Fresnel numbers \cite{paganin2006}, namely for sufficiently small sample-to-detector propagation distances $z_2$, Eq.~(\ref{eq:1}) can be simplified as follows:
\begin{eqnarray} \label{eq:2}
I_{z_1+z_2} (x,y) = I_{z_1} (x,y) -\frac{z_2}{k} \nabla_{\perp}\cdot[I_{z_1}(x,y)\nabla_{\perp}\varphi(x,y)].  \nonumber \\
\end{eqnarray}
Here $I_{z_1+z_2} (x,y)$ is the 2D intensity distribution after the wave-field propagates the distance $z_2$ from the object towards the detector.

Let us rewrite Eq.~(\ref{eq:2}) in a different form:
\begin{eqnarray}
\label{eq:18}
    \nonumber I_{z_1+z_2}({\bf{r}}_{\perp}) = \left[ I_{z_1}({\bf{r}}_{\perp}) -\frac{z_2}{k} \nabla_{\perp} I_{z_1}({\bf{r}}_{\perp}) \cdot \nabla_{\perp} \varphi({\bf{r}}_{\perp}) \right]  \\ \nonumber -\frac{z_2}{k} I_{z_1}({\bf{r}}_{\perp}) \nabla_{\perp}^2 \varphi({\bf{r}}_{\perp}) \\ \approx  I_{z_1}\left({\bf{r}}_{\perp}-\frac{z_2}{k}\nabla_{\perp}\varphi({\bf{r}}_{\perp})\right)-\frac{z_2}{k}I_{z_1}({\bf{r}}_{\perp})\nabla_{\perp}^2\varphi({\bf{r}}_{\perp}). ~~
\end{eqnarray}
In writing the final line of the above equation, we have combined two terms on the right-hand side and neglected higher order terms; note also that we have introduced the notation $(x,y)\equiv {\bf{r}}_{\perp}$.  Similar reasoning leads to:
\begin{eqnarray}
\label{eq:19}
    \nonumber {I}_{z_1+z_2}({\bf{r}}_{\perp}) & \approx & \tilde{I}_{z_1}\left({\bf{r}}_{\perp}-\frac{z_2}{k}\nabla_{\perp}\varphi_{\textrm{m}}({\bf{r}}_{\perp})-\frac{z_2}{k}\nabla_{\perp}\varphi_{\textrm{ob}}({\bf{r}}_{\perp})\right) \\ \nonumber & \times & I_{\textrm{ob}}\left({\bf{r}}_{\perp}-\frac{z_2}{k}\nabla_{\perp}\varphi_{\textrm{m}}({\bf{r}}_{\perp})-\frac{z_2}{k}\nabla_{\perp}\varphi_{\textrm{ob}}({\bf{r}}_{\perp})\right) \\ & \times & \left[ 1-\frac{z_2}{k}\nabla^2_{\perp}\varphi_{\textrm{m}}({\bf{r}}_{\perp}) -\frac{z_2}{k}\nabla^2_{\perp}\varphi_{\textrm{ob}}({\bf{r}}_{\perp})  \right].
\end{eqnarray}
Here $\tilde{I}_{z_1}(x,y)$ is the 2D intensity distribution just before the object (see Fig.~\ref{fig:ExperimentalSetUp}). As we consider a general case of the object, we may write: 
\begin{eqnarray}
\label{eq:20}
    {I}_{z_1}({\bf{r}}_{\perp}) & \approx & \tilde{I}_{z_1}({\bf{r}}_{\perp}){I}_{\textrm{ob}}({\bf{r}}_{\perp}),
\end{eqnarray}
where ${I}_{\textrm{ob}}({\bf{r}}_{\perp})$ describes the attenuation caused by the object. We can also write an expression similar to Eq.~(\ref{eq:19}) for the case when there is no object in the beam:
\begin{eqnarray}
\label{eq:21}
    \nonumber \tilde{I}_{z_1+z_2}({\bf{r}}_{\perp}) & \approx & \tilde{I}_{z_1}\left({\bf{r}}_{\perp}-\frac{z_2}{k}\nabla_{\perp}\varphi_{\textrm{m}}({\bf{r}}_{\perp})\right) \\ & \times & \left[ 1-\frac{z_2}{k}\nabla^2_{\perp}\varphi_{\textrm{m}}({\bf{r}}_{\perp})  \right].
\end{eqnarray}

Equations (\ref{eq:19}) and (\ref{eq:21}) provide a general formalism describing speckle-based imaging. They also describe the intensity distributions in certain other mask (grating) based phase-contrast imaging techniques. This formalism is somewhat similar to that for combined analyzer-based and propagation-based phase-contrast imaging \cite{pavlov2004,coan2005,pavlov2005}, where the terms responsible for the shift depend on the gradient of the phase, and the terms in brackets describe free-space propagation effects. Because cross-correlation approaches to phase retrieval (see e.g., \citet{berujon2012}, \citet{morgan2012}, \citet{zdora2018} and references therein) look simply at sample-induced distortions of the speckle pattern, the propagation-based edge effects described by the terms in brackets are not incorporated. However, ignoring these terms may produce artifacts in the reconstructed images. For instance, the usual procedure to obtain the object attenuation term (uncompensated for the effects of free-space propagation), $\tilde{I}_{\textrm{ob}}({\bf{r}}_{\perp})$, is based on dividing (i) the registered intensity when both the mask and object are in the beam (corrected to mitigate the refraction shift caused by the object), by (ii) the intensity when only the mask is present (see e.g., \citet{Morgan2013} and \citet{Zhou2015}).  Thus we may write:
\begin{eqnarray}
\label{eq:22}
\nonumber\tilde{I}_{\textrm{ob}}({\bf{r}}_{\perp}) &=& \frac{I_{z_1+z_2}\left({\bf{r}}_{\perp}+\frac{z_2}{k}\nabla_{\perp}\varphi_{\textrm{ob}}({\bf{r}}_{\perp})\right)}{\tilde{I}_{z_1+z_2}({\bf{r}}_{\perp})} \quad\quad\quad\quad\quad \\ &\approx&{I}_{\textrm{ob}}({\bf{r}}_{\perp})\left [ 1-\frac{z_2}{k}\nabla^2_{\perp}\varphi_{\textrm{ob}}({\bf{r}}_{\perp})\right].
\end{eqnarray}

The uncompensated free-space propagation term in brackets in Eq.~(\ref{eq:22}) can cause fringes in the reconstructed attenuation images to be evident (see e.g.~Fig.~2f in \citet{Morgan2013} and Fig.~3b in \citet{Zhou2015}), which require removal via a post-retrieval correction if the attenuation image is to be quantitative \cite{Groenendijk2020}. However, if we consider a ratio of unshifted intensities, we obtain an expression similar to the standard TIE (cf.~Eq.~(\ref{eq:2})) having only functions relating to the object in its right-hand side:
\begin{eqnarray}
\label{eq:23}
\nonumber\frac{I_{z_1+z_2}\left({\bf{r}}_{\perp}\right)}{\tilde{I}_{z_1+z_2}({\bf{r}}_{\perp})}   \approx I_{\textrm{ob}} \left({\bf{r}}_{\perp}\right) -\frac{z_2}{k} \nabla_{\perp}\cdot[I_{\textrm{ob}}\left({\bf{r}}_{\perp}\right)\nabla_{\perp}\varphi_{\textrm{ob}}\left({\bf{r}}_{\perp}\right)].\\
\end{eqnarray}
The result shown in Eq.~(\ref{eq:23}) is obtained in the case of a general object. In Eqs.~(\ref{eq:22}) and (\ref{eq:23}) we have neglected terms containing the averaged scalar product of the gradient of a random field and the gradient of a slowly varying function. However, the solution of this differential equation significantly simplifies in the case of a monomorphous (e.g., single-material) object.

As we have assumed that the object is monomorphous, its complex index of refraction $n$ obeys
\begin{eqnarray}
\label{eq:3}
n=1-\Delta+i\beta=1-\gamma\beta+i\beta=1+\beta(i-\gamma),
\end{eqnarray}
where 
\begin{equation}
\gamma=\frac{\Delta}{\beta}.    
\end{equation}
Here, the real numbers $\Delta,\beta$ respectively denote the refractive index decrement and the imaginary (absorptive) part of the complex refractive index. The value of $\gamma$ is considered known from tables or can be experimentally adjusted by trial and error to match the sample composition and density (see e.g., \citet{delRio2011}).
With reference to Fig.~\ref{fig:ExperimentalSetUp}, we seek to reconstruct the projected thickness $t(x,y)$ of the thin single-material object, which defines both the object's phase shift
\begin{eqnarray}
\varphi_{\textrm{ob}}\left({\bf{r}}_{\perp}\right)=-k\Delta \, t\left({\bf{r}}_{\perp}\right)=-k\gamma\beta \, t\left({\bf{r}}_{\perp}\right) 
\end{eqnarray}
and the object's absorption term
\begin{eqnarray}
\label{eq:6}
I_{ob}\left({\bf{r}}_{\perp}\right)=\exp[-2k\beta \, t\left({\bf{r}}_{\perp}\right)].
\end{eqnarray}
Then Eq.~(\ref{eq:23}) can be rewritten in the following form:
\begin{eqnarray}
\label{eq:24}
\nonumber\frac{I_{z_1+z_2}\left({\bf{r}}_{\perp}\right)}{\tilde{I}_{z_1+z_2}({\bf{r}}_{\perp})} &\approx& \left[ 1-\frac{\gamma z_2}{2 k}\nabla^2_{\perp}\right]I_{\textrm{ob}}\left({\bf{r}}_{\perp}\right) \\&=& \left[1-\frac{\gamma z_2}{2 k}\nabla^2_{\perp}\right]\exp\left[-2k\beta \, t({\bf{r}}_{\perp})\right].
\end{eqnarray}
In the case of a weakly-attenuating-object approximation, which is relevant to the experimental data used in this paper, we may write $\exp(-2k\beta \, t({\bf{r}}_{\perp})) \approx 1-2k\beta \, t({\bf{r}}_{\perp})$ and hence obtain:
\begin{eqnarray}
\label{eq:25}
\frac{I_{z_1+z_2}\left({\bf{r}}_{\perp}\right)}{\tilde{I}_{z_1+z_2}({\bf{r}}_{\perp})}-1= \left[1-\frac{\gamma z_2}{2 k}\nabla^2_{\perp}\right]\left[-2k\beta \, t({\bf{r}}_{\perp})\right].
\end{eqnarray}
This is surprising insofar as the data function defined by the left-hand side of Eq.~(\ref{eq:25}) is proportional to the Laplacian $\nabla_{\perp}^2$ of the projected thickness, rather than to the components of its transverse first derivative. Finally, the projected thickness in the case of a weakly attenuating object is
\begin{eqnarray}
\label{eq:11}
t(x,y) = \quad\quad\quad\quad\quad\quad\quad\quad\quad\quad\quad\quad\quad\quad\quad\quad\quad\quad\quad\quad \\ \nonumber \frac{1}{\mu} \, \mathcal{F}^{-1}\left( \textrm{LFF} \left\{ \frac{\mathcal{F}\left[1-{I_{z_1+z_2}(x,y)}/{\tilde{I}_{z_1+z_2}(x,y)}\right]}{1+\pi\gamma z_2 \lambda (u^2+v^2)} \right\} \right). 
\end{eqnarray}
Here, the linear attenuation coefficient $\mu$ is given by
\begin{eqnarray}
\label{eq:12}
\mu=2k\beta, 
\end{eqnarray}
$\mathcal{F}$ denotes Fourier transformation with respect to $x$ and $y$, $(u,v)$ are Fourier coordinates dual to $(x,y)$, $\mathcal{F}^{-1}$ denotes inverse Fourier transformation with respect to $u$ and $v$, and LFF is a optional high-pass filter. 
The projected thickness in the case of a relatively highly attenuating monomorphous (HAM) object is:

\begin{eqnarray}
\label{eq:11a}
t_{\textrm{HAM}}(x,y) = \quad\quad\quad\quad\quad\quad\quad\quad\quad\quad\quad\quad\quad\quad\quad\quad\quad\quad\quad \\ \nonumber -\frac{1}{\mu} \, \log_e\mathcal{F}^{-1}\left( \textrm{LFF} \left\{ \frac{\mathcal{F}\left[{I_{z_1+z_2}(x,y)}/{\tilde{I}_{z_1+z_2}(x,y)}\right]}{1+\pi\gamma z_2 \lambda (u^2+v^2)} \right\} \right). 
\end{eqnarray}
If the single-material object is embedded in another material with complex refractive index $n=1-\Delta_{\textrm{emb}}+i\beta_{\textrm{emb}}$, the effective values of
\begin{eqnarray}
\label{eq:13}
\Delta_{\textrm{eff}}=\Delta-\Delta_{\textrm{emb}} {\textrm{ and }} \beta_{\textrm{eff}}=\beta-\beta_{\textrm{emb}}
\end{eqnarray}
should be used (cf.~Gureyev {\em et al}.~\cite{Gureyev2002}). 

We close this section by considering how the blurring effects of finite source size and detector-based smearing may be taken into account in the reconstruction. Following Subbarao {\em et al.}~\cite{Unsharp0}, Gureyev {\em et al.}~\cite{DeblurByDefocus} and Beltran {\em et al.}~\cite{beltran2018}, we begin by noting that the linear differential operator
\begin{equation}
\label{eq:14}
\mathcal{L}_{\sigma}=1+\frac{1}{2}\sigma^2\nabla_{\perp}^2
\end{equation}
may be used to approximate the operation of convolving an image with a normalized rotationally symmetric Gaussian point spread function having a standard deviation of $\sigma$. Thus $\mathcal{L}_{\sigma}$ is a diffusion-type linear operator that blurs over a transverse length scale of $\sigma$. It may therefore be used to approximately model the effects of image smearing due to both finite source size and detector-induced blur. Applying this blurring operator $\mathcal{L}_{\sigma}$ to the right side of Eq.~(\ref{eq:25}), expanding the resulting operator product, and then discarding the term containing the bi-Laplacian $\nabla_{\perp}^2\nabla_{\perp}^2$, gives
\begin{eqnarray}
\label{eq:15}
\frac{I_{z_1+z_2}(x,y)}{\tilde{I}_{z_1+z_2}(x,y)}-1 \quad\quad\quad\quad\quad\quad\quad\quad\quad\quad\quad\quad\quad \\ \nonumber = \left[ 1- \frac{1}{2}\left(\frac{\gamma z_2}{k}-\sigma^2\right)\nabla_{\perp}^2 \right] [-2k\beta \, t(x,y)].
\end{eqnarray}
Since this differs from Eq.~(\ref{eq:25}) only through a different coefficient for the Laplacian, the corresponding solution in Eq.~(\ref{eq:11}) still holds, upon making a substitution similar to that given by Beltran {\em et al.}~in a different context \cite{beltran2018}:
\begin{eqnarray}
\label{eq:16}
\gamma z_2 \longrightarrow \gamma z_2 - k \sigma^2, \quad \gamma z_2 \ge k \sigma ^2.
\end{eqnarray}
Interestingly, if the blurring is such that
\begin{equation}
\label{eq:17}
\gamma z_2 = k \sigma ^2,    
\end{equation}
the effects of blurring exactly compensate the phase contrast, hence in this particular case the denominator in Eq.~(\ref{eq:11}) may be replaced with unity. This phenomenon is directly analogous to that discussed by Gureyev {\em et al.}~\cite{DeblurByDefocus} in the context of propagation-based x-ray phase contrast, whereby the sharpening phase-contrast effects of coherent propagation may be exactly balanced against the blurring effects due to source-size and detector-induced smearing \cite{Mittone2017}.

We emphasize that the preceding calculations, which are applied in the present context of our implicit approach to x-ray speckle imaging which does not rely on the use of correlation windows, are more broadly applicable.  In particular, the above calculations and their evident Fokker--Planck-type generalizations \cite{Risken1989,MorganPaganin2019,PaganinMorgan2019} may also be employed in contexts such as the ``X-ray Speckle-Vector Tracking'' (XSVT) formalism of \citeauthor{Berujon2015c} \cite{Berujon2015c}, together with the ``Unified Modulated Pattern Analysis'' (UMPA)  formalism of \citeauthor{Zdora2017} \cite{Zdora2017}.  Thus, while only one particular aspect of our formalism is implemented in the present paper, we believe it to be applicable to and indeed consistent with the two major formalisms  currently employed for x-ray speckle tracking, namely XSVT and UMPA.

\section{\label{sec:experiment1}Experiments}

Two distinct x-ray experiments, one of which employed monochromated radiation and the other of which employed broad-band polychromatic radiation, were conducted at the ID17 Biomedical Beamline of the European Synchrotron (ESRF).  

The first experiment aimed at testing the method under ideal conditions, i.e.~with a quasi monochromatic collimated beam. A double Si(111) crystal system in a Laue--Laue configuration was used to select x-ray photons from the synchrotron light source with an energy $E=52$ keV and a monochromaticity defined by the relative energy spread $\Delta E/E \simeq 10^{-4}$. Collimation was defined by the natural divergence of the 21-pole wiggler source which was less than $\sim$1~mrad horizontally and $\sim$0.1~mrad vertically. The photons had first to pass through the speckle-generator membrane before traversing the sample located $z_1=900$~mm further away as sketched in Fig.~\ref{fig:ExperimentalSetUp}. Eventually, the x rays impinged onto a detector placed $z_2 = 12$~m further downstream of the sample. Our detector consisted of an indirect detection system, based on a scientific CMOS camera coupled to magnifying optics imaging a scintillator screen made of a 60 $\mu$m thick gadolinium oxysulfide sheet as active layer. The resulting pixel size was $\simeq$ 6.1~$\mu$m. For this first experiment, a home made phantom composed of nylon wires of 150~$\mu$m in diameter was imaged to assess the accuracy of the method.  For these nylon wires at 52 keV we used the following parameters for the complex refractive index:  $\beta = 0.2809\times 10^{-11}; \Delta= 0.96842 \times 10^{-7}$ \cite{StepanovWebsite}.

To investigate further the potential applicability of the methodology to more polychromatic beams, another experiment was carried out using a pink beam tomographic configuration. After filtration of the white synchrotron beam by 0.5~mm of Al and 0.35~mm of Cu, the resulting utilized beam exhibited a peak with a mean energy of $E=37.3$~keV and a spectral bandwidth of approximately 20 keV ($\Delta E/E \simeq 0.5$). This second experiment aimed at testing the methodology by imaging in 3D a healthy mouse knee. Such a sample is composed of different tissue types with strongly varying refraction and absorption indices, and may therefore test under real-life conditions the efficacy of the algorithm, even for samples that are not truly weakly absorbing single-material objects. One can recall for instance the method of \citeauthor{Paganin_single_material_2002} \cite{Paganin_single_material_2002} which eventually proved efficient as well for a much wider range of samples than the class initially set by the strict conditions articulated in that paper. In accordance with Directive 2010/63/EU, the experiments were performed in a mutually agreed animal facility (C3851610006) which was evaluated and authorized by an Ethical Committee for Animal Welfare (APAFIS \#13792-201802261434542 v3). Mice were monitored for 10 weeks for any obvious locomotor disability by visual observation and Rotarod testing (Bioseb, Vitrolles, France) before being sacrificed. The imaged mouse knee was fixed with a 4\% formaldehyde solution for 48h after its removal and later embedded in a phosphate buffer solution with 2\% agarose.  For these mouse-knee data the value for $\gamma$ used for the reconstruction was $\gamma = 738$, which was obtained from tabulated values of $\Delta=2.98 \times 10^{-7}$ and $\beta=4.04\times 10^{-10}$ known for bones \cite{delRio2011}.

The detector system used to image the mouse knee was equivalent to that which was employed for the first experiment, but this time configured with more strongly magnifying optics which provided a resulting pixel size of $\simeq$ 3 $\mu$m. Overall, the setup layout was also equivalent to the one employed during the first experiment and sketched in Fig. \ref{fig:ExperimentalSetUp}, this time with the sample-to-detector and the membrane-to-sample distances respectively set to $z_2=4$~m and $z_1=1$~m. The tomography data consisted of 3000 projections collected periodically during a 360 degree scan of the sample. The center of rotation was transversely off-centered by 200 pixels to operate a so-called half acquisition tomography scan which is used to extend the 3D field of view. The phase images were calculated for each projection using Eq.~(\ref{eq:11}). The 3D computed-tomography (CT) reconstruction was performed using a filtered back-projection algorithm whose implementation is described in \citet{mirone2014}. For the sake of comparison the same raw data were processed using our earlier algorithm \cite{PaganinLabrietBrunBerujon2018} and the recovered phase subsequently used for volume reconstruction.

\section{\label{sec:experiment2}Results}

\begin{figure}
\includegraphics[trim=0 0 0 0,clip,width=0.48\textwidth,scale=0.2]{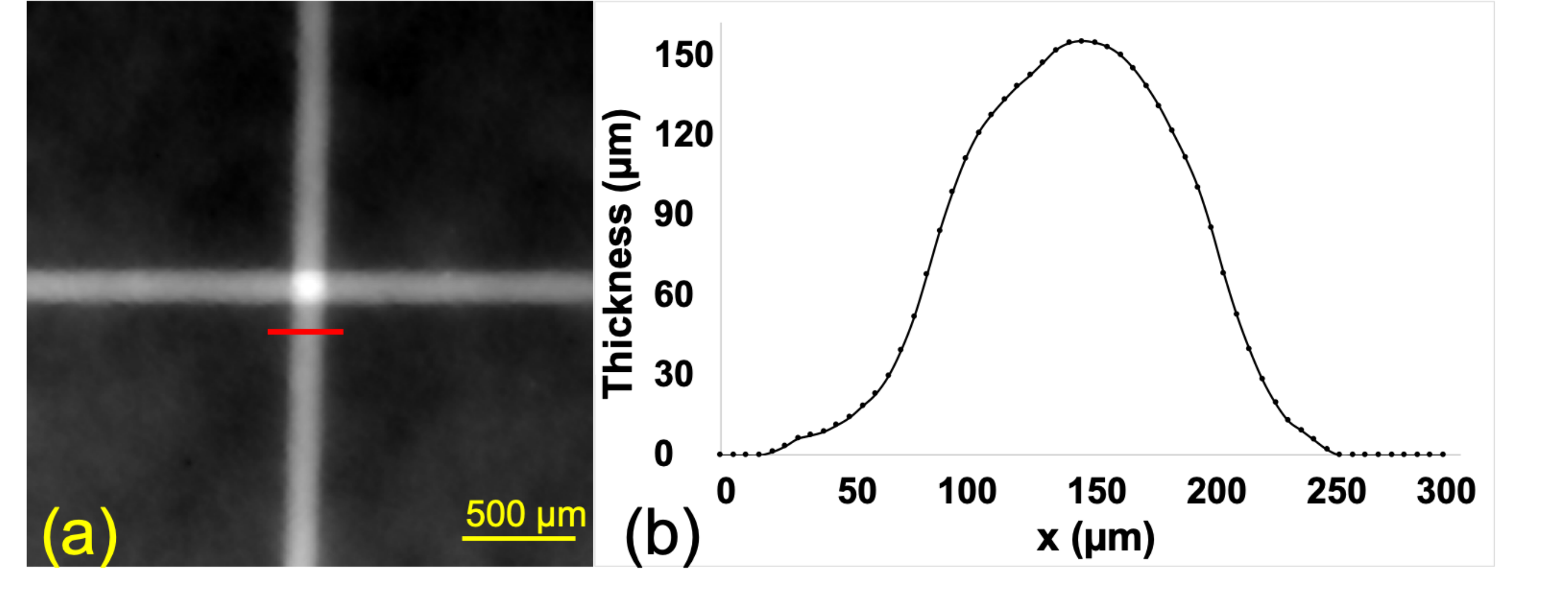}
\caption{Application of the method to monochromated x-ray radiation, with energy $E=52$ keV and energy spread $\Delta E / E \simeq 10^{-4}$. X-ray image of the nylon wire (a) and a profile of the calculated thickness (b) corresponding to the red line. }
\label{fig:wire}
\end{figure}

Figure \ref{fig:wire} presents the results of the calculated thickness on the nylon wire phantom, obtained using the monochromated beam with mean energy 52 keV and energy spread $\Delta E / E \simeq 10^{-4}$. The retrieved thickness is in good agreement with the nominal value of 150 $\mu$m wire diameter, showing the quantitativeness of our approach. In this case the LFF used in Eq.~(\ref{eq:11}) was a Gaussian filter, applied in Fourier space to the zero-padded image, with standard deviation $\sigma' = 7.5\delta_f$. Here $\delta_f$ is the step size of the variables \textit{u} and \textit{v} in the Fourier space. The $\sigma$ used in the modified form of Eq.~(\ref{eq:11}) as described in Eq.~(\ref{eq:16}) to obtain Fig.~\ref{fig:wire} is $150~\mu m$. The blurring effect, implied by Eq.~(\ref{eq:16}), causes the broadening of the wire's reconstructed line profile as evident from Fig.~\ref{fig:wire}.

\begin{figure*}
\includegraphics[trim=0 0 0 0,clip,width=0.88\textwidth]{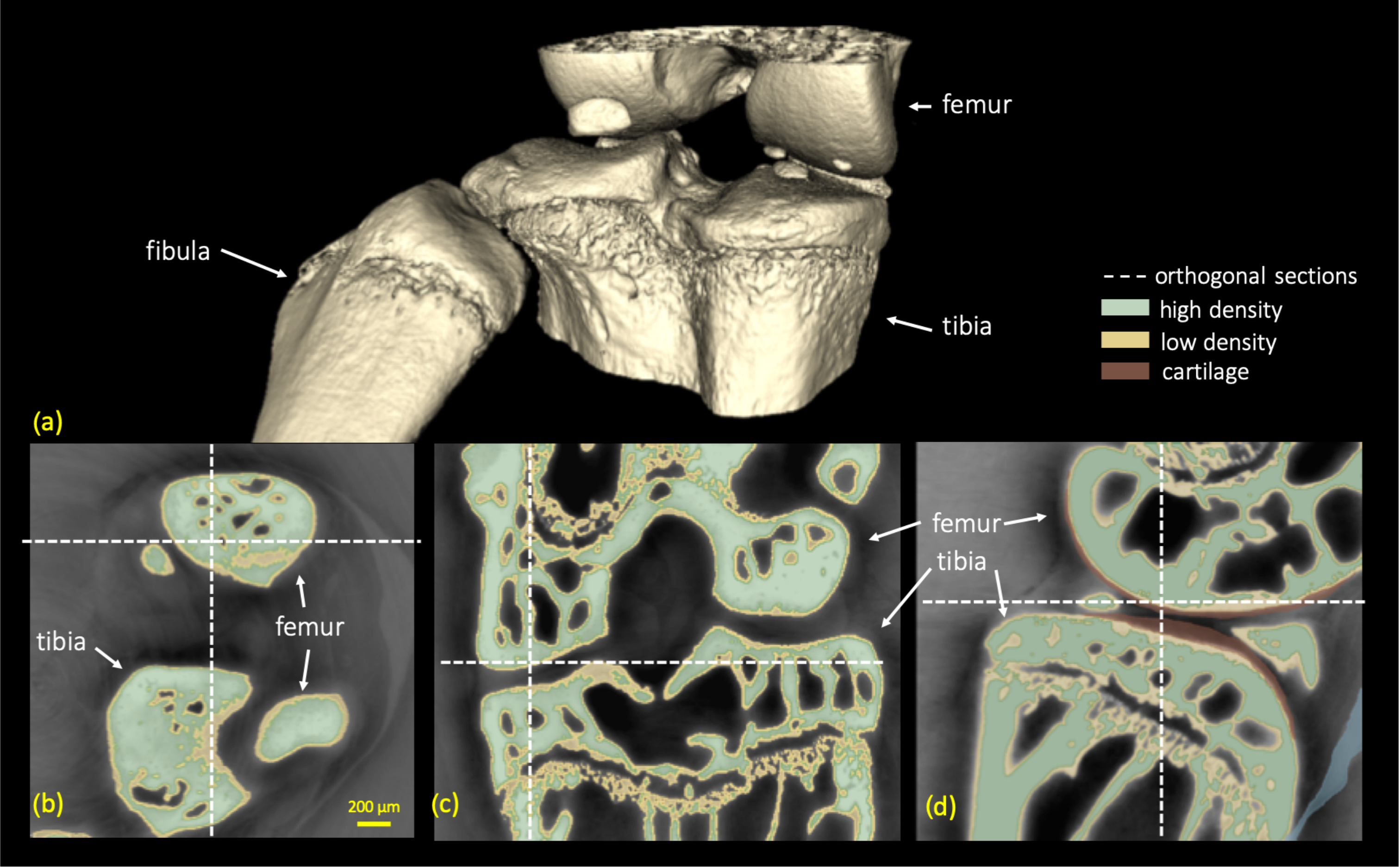}
\caption{Application of the method to polychromatic x-ray radiation, with mean energy $E=37.3$ keV and energy spread $\Delta E / E \simeq 0.5$. (a) 3D volume rendering of the relative density of the mouse knee. (b-d) Longitudinal cuts of the relative density. The sensitivity provided by the method permits a clear segmentation of the different tissues: ligament, cartilage and bones.}
\label{fig:3D_VR}
\end{figure*}

\begin{figure*}
\includegraphics[trim=0 0 0 0,clip,width=0.88\textwidth]{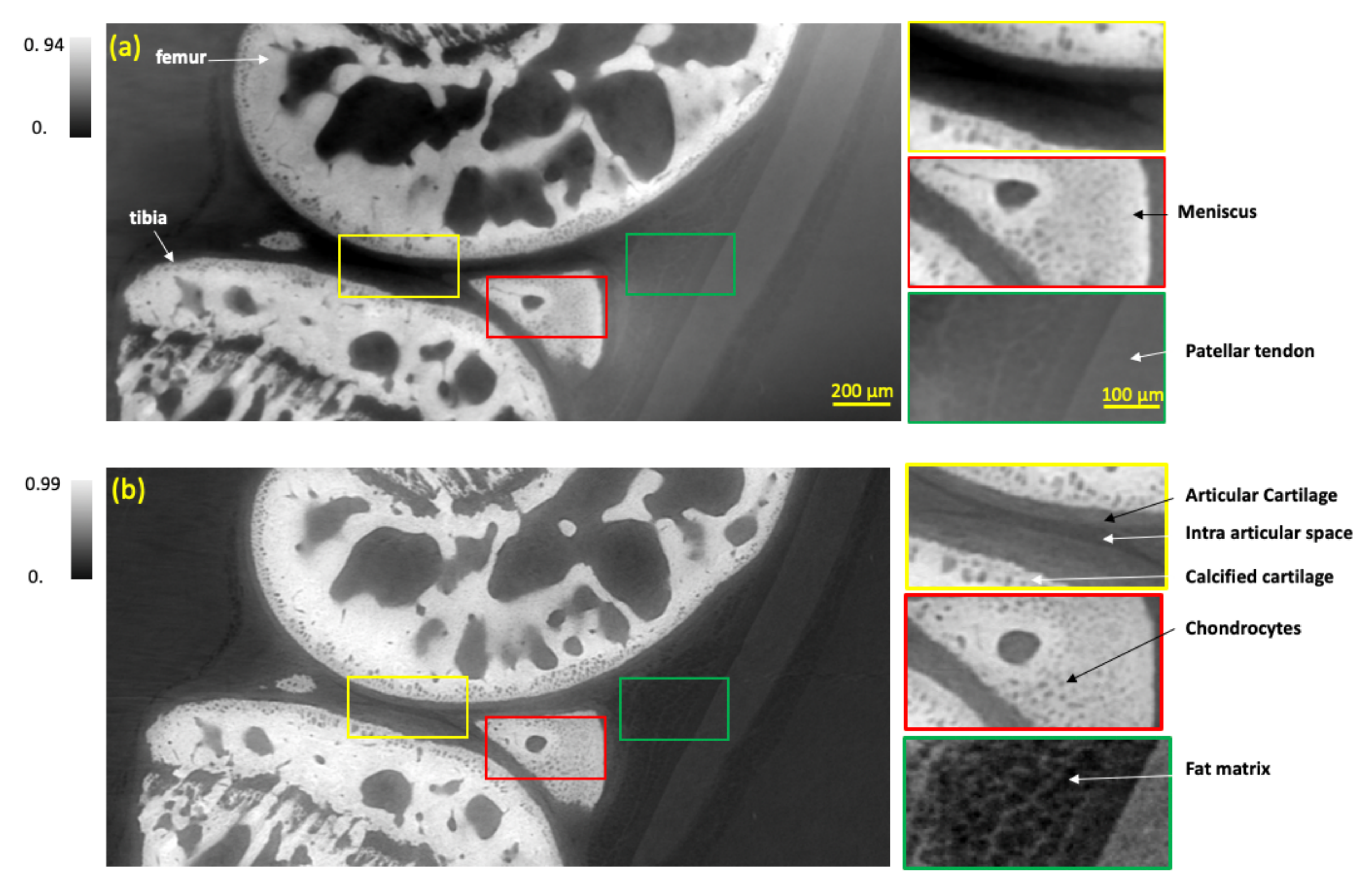}
\caption{Retrieved maps of (a) the function proportional to $\Delta \times k $ shown in arbitrary units and (b) the relative density of the mouse knee. The top row (a) presents results obtained using the method described in \citeauthor{PaganinLabrietBrunBerujon2018} \cite{PaganinLabrietBrunBerujon2018} while the bottom row, (b) shows the same knee part but reconstructed with the present method. The zoomed inset locations of the second column are marked with color squares on the sagittal slices.  Mean energy $E=37.3$ keV, energy spread $\Delta E / E \simeq 0.5$.}
\label{fig:slices}
\end{figure*}

Results of the tomography experiments, obtained using the broad-band polychromatic beam with mean energy $E=37.3$ keV and energy spread $\Delta E / E \simeq 0.5$, are presented in Figs.~\ref{fig:3D_VR} and \ref{fig:slices}. Figure~\ref{fig:3D_VR}(a) is a 3D volume rendering of the relative (to the reference material) density of the mouse knee and the subsets are 3$\mu$m thick axial (b), coronal (c) and sagittal (d) plane sections with anatomical annotations provided. The sensitivity of the technique permitted a clear differentiation of the various tissues composing the articulation. Although the hypothesis of a single-material object is not fulfilled here, one can nevertheless easily differentiate the hard tissues such as the bones from the softer tissue, e.g.~the ligaments. Moreover, tiny microcalcifications were also made visible with the technique. These microfeatures are usually investigated using conventional preclinical imaging modalities, but they nevertheless remain very difficult to detect. Indeed, the more traditional imaging modalities still encounter strong limitations in the detection of early cartilage and bony changes despite the crucial roles they play in the development of new therapeutic options \cite{bravin2012}. Such a feature of our current method can therefore occupy a place of value within the suite of x-ray phase contrast imaging techniques.

Figure~\ref{fig:slices} presents for comparison a sagittal thin slice from two 3D reconstructions obtained from the same raw data but using (a) the method of \citeauthor{PaganinLabrietBrunBerujon2018} \cite{PaganinLabrietBrunBerujon2018} to obtain the function proportional to $\Delta \times k $ of the mouse knee and (b) the herein presented method (without using the LFF) to obtain the relative density of the mouse knee. Insets of certain regions are presented in the right column of the figures. 

The yellow square in Fig.~\ref{fig:slices} marks a joint space region where the boundaries of the different tissues (bone, calcified cartilage, articular cartilage and the intra-articular space) are made visible thanks to the present method while most features remain barely distinguishable with the previous method of \citeauthor{PaganinLabrietBrunBerujon2018} \cite{PaganinLabrietBrunBerujon2018}. The method developed in the present paper allowed us in that case to measure articular cartilage thicknesses, which is of tremendous importance when evaluating for instance osteoarthritis. Such results are indicators of the present method being better able to handle the large difference between the bone ($\Delta=2.98 \times 10^{-7}$ and $\beta=4.04\times 10^{-10}$) and cartilage ($\Delta=1.8 \times 10^{-7}$ and $\beta=9.03\times 10^{-11}$) refractive index decrements \cite{delRio2011}. The red insets are located on one meniscus where the number of detectable chondrocytes (cartilage cells depicted as black round structures) appears higher than in the top image. This higher resolution in the hard tissue is confirmed also with the green inset covering a fat-matrix region. Therein, the honeycomb-like matrix is better resolved with the current method than with the former one.

\section{\label{sec:discussion}Discussion}

The x-ray speckle-tracking method may be viewed as a form of x-ray Hartmann--Shack sensor \cite{MayoSexton2004,Berto2017}, which employs distortions in random rather than regular arrays of reference intensity maxima, to measure the phase of an x-ray wave front. Moreover, in contrast to typical approaches to both Hartmann--Shack sensing and speckle tracking---in an x-ray setting and more broadly---our method implicitly rather than explicitly tracks speckles \cite{PaganinLabrietBrunBerujon2018}.  As noted earlier, this potentially increases the intrinsic resolution of the reconstruction since correlation windows, which must be at least a few pixels by a few pixels in size, are not needed. Further, our method is computationally simple and rapid to implement when compared to approaches based on optimizing error metrics in high-dimensional parameter spaces: regarding the latter, see references contained in the recent review by Zdora \cite{zdora2018}. We also note some parallels with earlier work in the visible-light regime, by Massig \cite{Massig1,Massig2} and Perciante {\em et al.} \cite{Perciante}.

Interestingly, the Fourier filtration in Eq.~(\ref{eq:11a}) is mathematically identical in form to the method of \citeauthor{Paganin_single_material_2002} \cite{Paganin_single_material_2002} (PM) for propagation-based x-ray phase contrast imaging of a single-material object. Two conclusions immediately follow. (i) Observe an interesting mathematical similarity, in the sense remarked upon earlier in this paragraph, between propagation-based phase contrast methods (which use no mask and do not rely on speckle) and speckle-tracking phase contrast methods (which use a mask to generate illuminating speckles). These two classes of method have typically been treated as distinct hitherto. (ii) The previously noted similarity opens the possibility for working in very low-dose environments, on account of the PM's very high robustness with respect to noise. For example, for the PM, 2D imaging with only one imaging quantum per pixel is possible \cite{Clark2019}. Several papers \cite{beltran2010,beltran2011,SNRboost3,SNRboost4,SNRboost5,gureyev2017unreasonable} show that signal-to-noise ratio (SNR) boosts well in excess of 100 are possible with the PM, which may be traded against dose (acquisition time) reductions of four orders of magnitude or more. This is of particular importance for contexts such as (a) dynamic samples and real-time x-ray imaging; (b) radiation-sensitive samples; (c) low-flux sources such as laboratory x-ray sources; (d) high-throughput scenarios such as industrial-product quality-control inspection. 

While the reconstructions in the present paper were of sufficient quality to not necessitate further refinement, it is also important to note in very-low-flux contexts that our algorithm could be used as an initial estimate to seed iterative refinement engines, thus enabling improved reconstructions that explicitly leverage relevant {\em a priori} knowledge such as real-space sparsity, Fourier-space sparsity etc. This is aligned with the usual trade-off that can be made where appropriate, in which an increase in suitable {\em a priori} knowledge may be traded off against the amount of data that one obtains (and therefore the dose incurred): more {\em a priori} knowledge allows one to make use of less data at the price of reduced generality, whereas less {\em a priori} knowledge requires more data to achieve reconstructions that are valid under more-general conditions. Recent achievements in phase-retrieval utilizing both compressive sensing and neural networks \cite{kemp2018propagation,rivenson2018phase,jiang2018solving} are particularly relevant in this regard, albeit beyond the scope of the present paper.

It is also interesting to note that our method overcomes a key issue for TIE imaging for x rays {\em without} the assumption of a single-material object, namely the need to acquire images in parallel planes for different $z$ values along the optical axis in order to retrieve quantitative sample information. While methods exist for single-plane x-ray TIE imaging, based e.g.~on measurements taken at different energies \cite{Li:18,Gursoy:13}, the method developed in the present paper is considerably more stable with respect to noise, again on account of its similarity in mathematical form to the PM. 

We close this discussion by noting that source code used for all analyses in this paper is freely available at:

\url{https://github.com/labrieth/spytlab}.

\section{\label{sec:conclusion}Conclusion}

We have developed and tested a means for x-ray speckle-tracking phase imaging, which extends our previous work approaching the problem using the concept of geometric flow \cite{PaganinLabrietBrunBerujon2018}. The method decodes the deformations in a single speckle field, created by a mask and subsequently deformed by an object, by applying transport-of-intensity concepts to this speckle-field scenario and then solving the inverse problem associated with the resulting equations. The ability to work with a single exposure, once the initial reference-speckle image has been taken in the absence of a sample, allows the method to be applied to time-dependent data. The stability with respect to noise of the method, which may be traded off against reduced dose and therefore reduced data acquisition time, means that our single-image method may be applicable to time-dependent imaging of dynamic samples. This ability to work with dynamic samples might be improved still further by the fact that coherence is not used explicitly in our method, which essentially relies on geometric optics alone, with associated very relaxed temporal coherence requirements. This enables it to be used with broad-band polychromatic beams, whose higher flux relative to monochromated beams may enable frame rates to be increased still further, in the context of imaging time-dependent samples. More generally, we believe that our method for implicitly tracking x-ray speckles has the capacity to make x-ray speckle tracking into a significantly more practical, dose-efficient, stable, rapid technique. Lastly, given the very recent translation of our previously published method for implicit speckle tracking \cite{PaganinLabrietBrunBerujon2018} to visible-light microscopy \cite{Lu2019}, we anticipate that the method developed in the present paper may also be of utility in the visible-light domain.

\acknowledgements

We acknowledge useful discussions with Claudio Ferrero, Timur Gureyev, Andrew Kingston, Marcus Kitchen, Sheridan Mayo, Kaye Morgan, and Gary Ruben. Financial support from the Experiment Division of the European Synchrotron Radiation Facility (ESRF) for DMP to visit in early 2018 is gratefully acknowledged. EB acknowledges support from LabEx PRIMES (ANR-11-LABX-0063/ANR-11-IDEX-0007).

\bibliography{refbibtek}

\begin{thebibliography}{51}%
\makeatletter
\providecommand \@ifxundefined [1]{%
 \@ifx{#1\undefined}
}%
\providecommand \@ifnum [1]{%
 \ifnum #1\expandafter \@firstoftwo
 \else \expandafter \@secondoftwo
 \fi
}%
\providecommand \@ifx [1]{%
 \ifx #1\expandafter \@firstoftwo
 \else \expandafter \@secondoftwo
 \fi
}%
\providecommand \natexlab [1]{#1}%
\providecommand \enquote  [1]{``#1''}%
\providecommand \bibnamefont  [1]{#1}%
\providecommand \bibfnamefont [1]{#1}%
\providecommand \citenamefont [1]{#1}%
\providecommand \href@noop [0]{\@secondoftwo}%
\providecommand \href [0]{\begingroup \@sanitize@url \@href}%
\providecommand \@href[1]{\@@startlink{#1}\@@href}%
\providecommand \@@href[1]{\endgroup#1\@@endlink}%
\providecommand \@sanitize@url [0]{\catcode `\\12\catcode `\$12\catcode
  `\&12\catcode `\#12\catcode `\^12\catcode `\_12\catcode `\%12\relax}%
\providecommand \@@startlink[1]{}%
\providecommand \@@endlink[0]{}%
\providecommand \url  [0]{\begingroup\@sanitize@url \@url }%
\providecommand \@url [1]{\endgroup\@href {#1}{\urlprefix }}%
\providecommand \urlprefix  [0]{URL }%
\providecommand \Eprint [0]{\href }%
\providecommand \doibase [0]{http://dx.doi.org/}%
\providecommand \selectlanguage [0]{\@gobble}%
\providecommand \bibinfo  [0]{\@secondoftwo}%
\providecommand \bibfield  [0]{\@secondoftwo}%
\providecommand \translation [1]{[#1]}%
\providecommand \BibitemOpen [0]{}%
\providecommand \bibitemStop [0]{}%
\providecommand \bibitemNoStop [0]{.\EOS\space}%
\providecommand \EOS [0]{\spacefactor3000\relax}%
\providecommand \BibitemShut  [1]{\csname bibitem#1\endcsname}%
\let\auto@bib@innerbib\@empty
\bibitem [{\citenamefont {Paganin}(2006)}]{paganin2006}%
  \BibitemOpen
  \bibfield  {author} {\bibinfo {author} {\bibfnamefont {D.~M.}\ \bibnamefont
  {Paganin}},\ }\href@noop {} {\emph {\bibinfo {title} {Coherent X-Ray
  Optics}}}\ (\bibinfo  {publisher} {Oxford University Press},\ \bibinfo
  {address} {Oxford},\ \bibinfo {year} {2006})\BibitemShut {NoStop}%
\bibitem [{\citenamefont {Bravin}\ \emph {et~al.}(2013)\citenamefont {Bravin},
  \citenamefont {Coan},\ and\ \citenamefont {Suortti}}]{bravin2012}%
  \BibitemOpen
  \bibfield  {author} {\bibinfo {author} {\bibfnamefont {A.}~\bibnamefont
  {Bravin}}, \bibinfo {author} {\bibfnamefont {P.}~\bibnamefont {Coan}}, \ and\
  \bibinfo {author} {\bibfnamefont {P.}~\bibnamefont {Suortti}},\ }\bibfield
  {title} {\enquote {\bibinfo {title} {{X-ray phase-contrast imaging: from
  pre-clinical applications towards clinics}},}\ }\href {\doibase
  10.1088/0031-9155/58/1/R1} {\bibfield  {journal} {\bibinfo  {journal} {Phys.
  Med. Biol.}\ }\textbf {\bibinfo {volume} {58}},\ \bibinfo {pages} {R1--R35}
  (\bibinfo {year} {2013})}\BibitemShut {NoStop}%
\bibitem [{\citenamefont {Berujon}\ \emph {et~al.}(2012)\citenamefont
  {Berujon}, \citenamefont {Ziegler}, \citenamefont {Cerbino},\ and\
  \citenamefont {Peverini}}]{berujon2012}%
  \BibitemOpen
  \bibfield  {author} {\bibinfo {author} {\bibfnamefont {S.}~\bibnamefont
  {Berujon}}, \bibinfo {author} {\bibfnamefont {E.}~\bibnamefont {Ziegler}},
  \bibinfo {author} {\bibfnamefont {R.}~\bibnamefont {Cerbino}}, \ and\
  \bibinfo {author} {\bibfnamefont {L.}~\bibnamefont {Peverini}},\ }\bibfield
  {title} {\enquote {\bibinfo {title} {Two-dimensional x-ray beam phase
  sensing},}\ }\href@noop {} {\bibfield  {journal} {\bibinfo  {journal} {Phys.
  Rev. Lett.}\ }\textbf {\bibinfo {volume} {108}},\ \bibinfo {pages} {158102}
  (\bibinfo {year} {2012})}\BibitemShut {NoStop}%
\bibitem [{\citenamefont {Morgan}\ \emph {et~al.}(2012)\citenamefont {Morgan},
  \citenamefont {Paganin},\ and\ \citenamefont {Siu}}]{morgan2012}%
  \BibitemOpen
  \bibfield  {author} {\bibinfo {author} {\bibfnamefont {K.~S.}\ \bibnamefont
  {Morgan}}, \bibinfo {author} {\bibfnamefont {D.~M.}\ \bibnamefont {Paganin}},
  \ and\ \bibinfo {author} {\bibfnamefont {K.~K.~W.}\ \bibnamefont {Siu}},\
  }\bibfield  {title} {\enquote {\bibinfo {title} {X-ray phase imaging with a
  paper analyzer},}\ }\href@noop {} {\bibfield  {journal} {\bibinfo  {journal}
  {Appl. Phys. Lett.}\ }\textbf {\bibinfo {volume} {100}},\ \bibinfo {pages}
  {124102} (\bibinfo {year} {2012})}\BibitemShut {NoStop}%
\bibitem [{\citenamefont {Zdora}(2018)}]{zdora2018}%
  \BibitemOpen
  \bibfield  {author} {\bibinfo {author} {\bibfnamefont {M.-C.}\ \bibnamefont
  {Zdora}},\ }\bibfield  {title} {\enquote {\bibinfo {title} {State of the art
  of x-ray speckle-based phase-contrast and dark-field imaging},}\ }\href@noop
  {} {\bibfield  {journal} {\bibinfo  {journal} {J. Imaging}\ }\textbf
  {\bibinfo {volume} {4}},\ \bibinfo {pages} {60} (\bibinfo {year}
  {2018})}\BibitemShut {NoStop}%
\bibitem [{\citenamefont {Paganin}\ \emph {et~al.}(2018)\citenamefont
  {Paganin}, \citenamefont {Labriet}, \citenamefont {Brun},\ and\ \citenamefont
  {Berujon}}]{PaganinLabrietBrunBerujon2018}%
  \BibitemOpen
  \bibfield  {author} {\bibinfo {author} {\bibfnamefont {D.~M.}\ \bibnamefont
  {Paganin}}, \bibinfo {author} {\bibfnamefont {H.}~\bibnamefont {Labriet}},
  \bibinfo {author} {\bibfnamefont {E.}~\bibnamefont {Brun}}, \ and\ \bibinfo
  {author} {\bibfnamefont {S.}~\bibnamefont {Berujon}},\ }\bibfield  {title}
  {\enquote {\bibinfo {title} {Single-image geometric-flow x-ray speckle
  tracking},}\ }\href
  {https://journals.aps.org/pra/abstract/10.1103/PhysRevA.98.053813} {\bibfield
   {journal} {\bibinfo  {journal} {Phys. Rev. A}\ }\textbf {\bibinfo {volume}
  {98}},\ \bibinfo {pages} {053813} (\bibinfo {year} {2018})}\BibitemShut
  {NoStop}%
\bibitem [{\citenamefont {Teague}(1983)}]{teague1983}%
  \BibitemOpen
  \bibfield  {author} {\bibinfo {author} {\bibfnamefont {M.~R.}\ \bibnamefont
  {Teague}},\ }\bibfield  {title} {\enquote {\bibinfo {title} {Deterministic
  phase retrieval: a {G}reen's function solution},}\ }\href
  {http://www.opticsinfobase.org/abstract.cfm?URI=josa-73-11-1434} {\bibfield
  {journal} {\bibinfo  {journal} {J. Opt. Soc. Am.}\ }\textbf {\bibinfo
  {volume} {73}},\ \bibinfo {pages} {1434--1441} (\bibinfo {year}
  {1983})}\BibitemShut {NoStop}%
\bibitem [{\citenamefont {Horn}\ and\ \citenamefont {Schunck}(1981)}]{OF_old1}%
  \BibitemOpen
  \bibfield  {author} {\bibinfo {author} {\bibfnamefont {B.~K.~P.}\
  \bibnamefont {Horn}}\ and\ \bibinfo {author} {\bibfnamefont {B.~G.}\
  \bibnamefont {Schunck}},\ }\bibfield  {title} {\enquote {\bibinfo {title}
  {Determining optical flow},}\ }\href@noop {} {\bibfield  {journal} {\bibinfo
  {journal} {Artif. Intell.}\ }\textbf {\bibinfo {volume} {17}},\ \bibinfo
  {pages} {185--203} (\bibinfo {year} {1981})}\BibitemShut {NoStop}%
\bibitem [{\citenamefont {Atcheson}\ \emph {et~al.}(2009)\citenamefont
  {Atcheson}, \citenamefont {Heidrich},\ and\ \citenamefont {Ihrke}}]{OF_old2}%
  \BibitemOpen
  \bibfield  {author} {\bibinfo {author} {\bibfnamefont {B.}~\bibnamefont
  {Atcheson}}, \bibinfo {author} {\bibfnamefont {W.}~\bibnamefont {Heidrich}},
  \ and\ \bibinfo {author} {\bibfnamefont {I.}~\bibnamefont {Ihrke}},\
  }\bibfield  {title} {\enquote {\bibinfo {title} {An evaluation of optical
  flow algorithms for background oriented schlieren imaging},}\ }\href@noop {}
  {\bibfield  {journal} {\bibinfo  {journal} {Exp. Fluids.}\ }\textbf {\bibinfo
  {volume} {46}},\ \bibinfo {pages} {467--476} (\bibinfo {year}
  {2009})}\BibitemShut {NoStop}%
\bibitem [{\citenamefont {Lu}\ \emph {et~al.}(2019)\citenamefont {Lu},
  \citenamefont {Sun}, \citenamefont {Zhang}, \citenamefont {Fan},
  \citenamefont {Chen},\ and\ \citenamefont {Zuo}}]{Lu2019}%
  \BibitemOpen
  \bibfield  {author} {\bibinfo {author} {\bibfnamefont {L.}~\bibnamefont
  {Lu}}, \bibinfo {author} {\bibfnamefont {J.}~\bibnamefont {Sun}}, \bibinfo
  {author} {\bibfnamefont {J.}~\bibnamefont {Zhang}}, \bibinfo {author}
  {\bibfnamefont {Y.}~\bibnamefont {Fan}}, \bibinfo {author} {\bibfnamefont
  {Q.}~\bibnamefont {Chen}}, \ and\ \bibinfo {author} {\bibfnamefont
  {C.}~\bibnamefont {Zuo}},\ }\bibfield  {title} {\enquote {\bibinfo {title}
  {Quantitative phase imaging camera with a weak diffuser},}\ }\href {\doibase
  10.3389/fphy.2019.00077} {\bibfield  {journal} {\bibinfo  {journal} {Front.
  Phys.}\ }\textbf {\bibinfo {volume} {7}},\ \bibinfo {pages} {77} (\bibinfo
  {year} {2019})}\BibitemShut {NoStop}%
\bibitem [{\citenamefont {Paganin}\ \emph {et~al.}(2004)\citenamefont
  {Paganin}, \citenamefont {Gureyev}, \citenamefont {Mayo}, \citenamefont
  {Stevenson}, \citenamefont {Nesterets},\ and\ \citenamefont
  {Wilkins}}]{Paganin2004}%
  \BibitemOpen
  \bibfield  {author} {\bibinfo {author} {\bibfnamefont {D.}~\bibnamefont
  {Paganin}}, \bibinfo {author} {\bibfnamefont {T.~E.}\ \bibnamefont
  {Gureyev}}, \bibinfo {author} {\bibfnamefont {S.~C.}\ \bibnamefont {Mayo}},
  \bibinfo {author} {\bibfnamefont {A.~W.}\ \bibnamefont {Stevenson}}, \bibinfo
  {author} {\bibfnamefont {{Ya}.~I.}\ \bibnamefont {Nesterets}}, \ and\
  \bibinfo {author} {\bibfnamefont {S.~W.}\ \bibnamefont {Wilkins}},\
  }\bibfield  {title} {\enquote {\bibinfo {title} {X-ray omni microscopy},}\
  }\href@noop {} {\bibfield  {journal} {\bibinfo  {journal} {J. Microsc.}\
  }\textbf {\bibinfo {volume} {214}},\ \bibinfo {pages} {315--327} (\bibinfo
  {year} {2004})}\BibitemShut {NoStop}%
\bibitem [{\citenamefont {Gureyev}\ \emph {et~al.}(2015)\citenamefont
  {Gureyev}, \citenamefont {Nesterets},\ and\ \citenamefont
  {Paganin}}]{GureyevMonomorphous2015}%
  \BibitemOpen
  \bibfield  {author} {\bibinfo {author} {\bibfnamefont {T.~E.}\ \bibnamefont
  {Gureyev}}, \bibinfo {author} {\bibfnamefont {Ya.~I.}\ \bibnamefont
  {Nesterets}}, \ and\ \bibinfo {author} {\bibfnamefont {D.~M.}\ \bibnamefont
  {Paganin}},\ }\bibfield  {title} {\enquote {\bibinfo {title} {Monomorphous
  decomposition method and its application for phase retrieval and
  phase-contrast tomography},}\ }\href@noop {} {\bibfield  {journal} {\bibinfo
  {journal} {Phys. Rev. A}\ }\textbf {\bibinfo {volume} {92}},\ \bibinfo
  {pages} {053860} (\bibinfo {year} {2015})}\BibitemShut {NoStop}%
\bibitem [{\citenamefont {Paganin}\ \emph {et~al.}(2002)\citenamefont
  {Paganin}, \citenamefont {Mayo}, \citenamefont {Gureyev}, \citenamefont
  {Miller},\ and\ \citenamefont {Wilkins}}]{Paganin_single_material_2002}%
  \BibitemOpen
  \bibfield  {author} {\bibinfo {author} {\bibfnamefont {D.}~\bibnamefont
  {Paganin}}, \bibinfo {author} {\bibfnamefont {S.~C.}\ \bibnamefont {Mayo}},
  \bibinfo {author} {\bibfnamefont {T.~E.}\ \bibnamefont {Gureyev}}, \bibinfo
  {author} {\bibfnamefont {P.~R.}\ \bibnamefont {Miller}}, \ and\ \bibinfo
  {author} {\bibfnamefont {S.~W.}\ \bibnamefont {Wilkins}},\ }\bibfield
  {title} {\enquote {\bibinfo {title} {Simultaneous phase and amplitude
  extraction from a single defocused image of a homogeneous object},}\
  }\href@noop {} {\bibfield  {journal} {\bibinfo  {journal} {J. Microsc.}\
  }\textbf {\bibinfo {volume} {206}},\ \bibinfo {pages} {33--40} (\bibinfo
  {year} {2002})}\BibitemShut {NoStop}%
\bibitem [{\citenamefont {Wu}\ \emph {et~al.}(2005)\citenamefont {Wu},
  \citenamefont {Liu},\ and\ \citenamefont {Yan}}]{Wu2005}%
  \BibitemOpen
  \bibfield  {author} {\bibinfo {author} {\bibfnamefont {X.}~\bibnamefont
  {Wu}}, \bibinfo {author} {\bibfnamefont {H.}~\bibnamefont {Liu}}, \ and\
  \bibinfo {author} {\bibfnamefont {A.}~\bibnamefont {Yan}},\ }\bibfield
  {title} {\enquote {\bibinfo {title} {X-ray phase-attenuation duality and
  phase retrieval},}\ }\href {\doibase 10.1364/OL.30.000379} {\bibfield
  {journal} {\bibinfo  {journal} {Opt. Lett.}\ }\textbf {\bibinfo {volume}
  {30}},\ \bibinfo {pages} {379--381} (\bibinfo {year} {2005})}\BibitemShut
  {NoStop}%
\bibitem [{Note1()}]{Note1}%
  \BibitemOpen
  \bibinfo {note} {As shall become clear later in the paper, the
  spatially-random intensity distributions that we consider are not
  fully-developed speckle. Much literature equates {\protect \em speckle} with
  {\protect \em fully developed speckle}, but we adopt a more broadly
  applicable usage of the term {\protect \em speckle} to refer to any
  spatially-random intensity distribution.}\BibitemShut {Stop}%
\bibitem [{\citenamefont {Pavlov}\ \emph {et~al.}(2004)\citenamefont {Pavlov},
  \citenamefont {Gureyev}, \citenamefont {Paganin}, \citenamefont {Nesterets},
  \citenamefont {Morgan},\ and\ \citenamefont {Lewis}}]{pavlov2004}%
  \BibitemOpen
  \bibfield  {author} {\bibinfo {author} {\bibfnamefont {K.~M.}\ \bibnamefont
  {Pavlov}}, \bibinfo {author} {\bibfnamefont {T.~E.}\ \bibnamefont {Gureyev}},
  \bibinfo {author} {\bibfnamefont {D.}~\bibnamefont {Paganin}}, \bibinfo
  {author} {\bibfnamefont {Ya.I.}\ \bibnamefont {Nesterets}}, \bibinfo {author}
  {\bibfnamefont {M.J.}\ \bibnamefont {Morgan}}, \ and\ \bibinfo {author}
  {\bibfnamefont {R.A.}\ \bibnamefont {Lewis}},\ }\bibfield  {title} {\enquote
  {\bibinfo {title} {Linear systems with slowly varying transfer functions and
  their application to x-ray phase-contrast imaging},}\ }\href@noop {}
  {\bibfield  {journal} {\bibinfo  {journal} {J. Phys. D: Appl. Phys}\ }\textbf
  {\bibinfo {volume} {37}},\ \bibinfo {pages} {2746--2750} (\bibinfo {year}
  {2004})}\BibitemShut {NoStop}%
\bibitem [{\citenamefont {Coan}\ \emph {et~al.}(2005)\citenamefont {Coan},
  \citenamefont {Pagot}, \citenamefont {Fiedler}, \citenamefont {Cloetens},
  \citenamefont {Baruchel},\ and\ \citenamefont {Bravin}}]{coan2005}%
  \BibitemOpen
  \bibfield  {author} {\bibinfo {author} {\bibfnamefont {P.}~\bibnamefont
  {Coan}}, \bibinfo {author} {\bibfnamefont {E.}~\bibnamefont {Pagot}},
  \bibinfo {author} {\bibfnamefont {S.}~\bibnamefont {Fiedler}}, \bibinfo
  {author} {\bibfnamefont {P.}~\bibnamefont {Cloetens}}, \bibinfo {author}
  {\bibfnamefont {J.}~\bibnamefont {Baruchel}}, \ and\ \bibinfo {author}
  {\bibfnamefont {A.}~\bibnamefont {Bravin}},\ }\bibfield  {title} {\enquote
  {\bibinfo {title} {Phase-contrast {X}-ray imaging combining free space
  propagation and {B}ragg diffraction},}\ }\href@noop {} {\bibfield  {journal}
  {\bibinfo  {journal} {J. Synchrotron Radiat.}\ }\textbf {\bibinfo {volume}
  {12}},\ \bibinfo {pages} {241--245} (\bibinfo {year} {2005})}\BibitemShut
  {NoStop}%
\bibitem [{\citenamefont {Pavlov}\ \emph {et~al.}(2005)\citenamefont {Pavlov},
  \citenamefont {Gureyev}, \citenamefont {Paganin}, \citenamefont {Nesterets},
  \citenamefont {Kitchen}, \citenamefont {Siu}, \citenamefont {Gillam},
  \citenamefont {Uesugi}, \citenamefont {Yagi}, \citenamefont {Morgan},\ and\
  \citenamefont {Lewis}}]{pavlov2005}%
  \BibitemOpen
  \bibfield  {author} {\bibinfo {author} {\bibfnamefont {K.~M.}\ \bibnamefont
  {Pavlov}}, \bibinfo {author} {\bibfnamefont {T.~E.}\ \bibnamefont {Gureyev}},
  \bibinfo {author} {\bibfnamefont {D.}~\bibnamefont {Paganin}}, \bibinfo
  {author} {\bibfnamefont {Ya.~I.}\ \bibnamefont {Nesterets}}, \bibinfo
  {author} {\bibfnamefont {M.~J.}\ \bibnamefont {Kitchen}}, \bibinfo {author}
  {\bibfnamefont {K.~K.~W.}\ \bibnamefont {Siu}}, \bibinfo {author}
  {\bibfnamefont {J.~E.}\ \bibnamefont {Gillam}}, \bibinfo {author}
  {\bibfnamefont {K.}~\bibnamefont {Uesugi}}, \bibinfo {author} {\bibfnamefont
  {N.}~\bibnamefont {Yagi}}, \bibinfo {author} {\bibfnamefont {M.~J.}\
  \bibnamefont {Morgan}}, \ and\ \bibinfo {author} {\bibfnamefont {R.~A.}\
  \bibnamefont {Lewis}},\ }\bibfield  {title} {\enquote {\bibinfo {title}
  {Unification of analyser-based and propagation-based x-ray phase-contrast
  imaging},}\ }\href@noop {} {\bibfield  {journal} {\bibinfo  {journal} {Nucl.
  Instr. Meth. A}\ }\textbf {\bibinfo {volume} {548}},\ \bibinfo {pages}
  {163--168} (\bibinfo {year} {2005})}\BibitemShut {NoStop}%
\bibitem [{\citenamefont {Morgan}\ \emph {et~al.}(2013)\citenamefont {Morgan},
  \citenamefont {Modregger}, \citenamefont {Irvine}, \citenamefont
  {Rutishauser}, \citenamefont {Guzenko}, \citenamefont {Stampanoni},\ and\
  \citenamefont {David}}]{Morgan2013}%
  \BibitemOpen
  \bibfield  {author} {\bibinfo {author} {\bibfnamefont {K.~S.}\ \bibnamefont
  {Morgan}}, \bibinfo {author} {\bibfnamefont {P.}~\bibnamefont {Modregger}},
  \bibinfo {author} {\bibfnamefont {S.~C.}\ \bibnamefont {Irvine}}, \bibinfo
  {author} {\bibfnamefont {S.}~\bibnamefont {Rutishauser}}, \bibinfo {author}
  {\bibfnamefont {V.~A.}\ \bibnamefont {Guzenko}}, \bibinfo {author}
  {\bibfnamefont {M.}~\bibnamefont {Stampanoni}}, \ and\ \bibinfo {author}
  {\bibfnamefont {C.}~\bibnamefont {David}},\ }\bibfield  {title} {\enquote
  {\bibinfo {title} {A sensitive x-ray phase contrast technique for rapid
  imaging using a single phase grid analyzer},}\ }\href@noop {} {\bibfield
  {journal} {\bibinfo  {journal} {Opt. Lett.}\ }\textbf {\bibinfo {volume}
  {38}},\ \bibinfo {pages} {4605--4608} (\bibinfo {year} {2013})}\BibitemShut
  {NoStop}%
\bibitem [{\citenamefont {Zhou}\ \emph {et~al.}(2015)\citenamefont {Zhou},
  \citenamefont {Zanette}, \citenamefont {Zdora}, \citenamefont
  {Lundstr{\"{o}}m}, \citenamefont {Larsson}, \citenamefont {Hertz},
  \citenamefont {Pfeiffer},\ and\ \citenamefont {Burvall}}]{Zhou2015}%
  \BibitemOpen
  \bibfield  {author} {\bibinfo {author} {\bibfnamefont {T.}~\bibnamefont
  {Zhou}}, \bibinfo {author} {\bibfnamefont {I.}~\bibnamefont {Zanette}},
  \bibinfo {author} {\bibfnamefont {M.-C.}\ \bibnamefont {Zdora}}, \bibinfo
  {author} {\bibfnamefont {U.}~\bibnamefont {Lundstr{\"{o}}m}}, \bibinfo
  {author} {\bibfnamefont {D.~H.}\ \bibnamefont {Larsson}}, \bibinfo {author}
  {\bibfnamefont {H.~M.}\ \bibnamefont {Hertz}}, \bibinfo {author}
  {\bibfnamefont {F.}~\bibnamefont {Pfeiffer}}, \ and\ \bibinfo {author}
  {\bibfnamefont {A.}~\bibnamefont {Burvall}},\ }\bibfield  {title} {\enquote
  {\bibinfo {title} {Speckle-based x-ray phase-contrast imaging with a
  laboratory source and the scanning technique},}\ }\href@noop {} {\bibfield
  {journal} {\bibinfo  {journal} {Opt. Lett.}\ }\textbf {\bibinfo {volume}
  {40}},\ \bibinfo {pages} {2822--2825} (\bibinfo {year} {2015})}\BibitemShut
  {NoStop}%
\bibitem [{\citenamefont {Groenendijk}\ \emph {et~al.}(2020)\citenamefont
  {Groenendijk}, \citenamefont {Schaff}, \citenamefont {Croton}, \citenamefont
  {Kitchen},\ and\ \citenamefont {Morgan}}]{Groenendijk2020}%
  \BibitemOpen
  \bibfield  {author} {\bibinfo {author} {\bibfnamefont {C.~F.}\ \bibnamefont
  {Groenendijk}}, \bibinfo {author} {\bibfnamefont {F.}~\bibnamefont {Schaff}},
  \bibinfo {author} {\bibfnamefont {L.~C.~P.}\ \bibnamefont {Croton}}, \bibinfo
  {author} {\bibfnamefont {M.~J.}\ \bibnamefont {Kitchen}}, \ and\ \bibinfo
  {author} {\bibfnamefont {K.~S.}\ \bibnamefont {Morgan}},\ }\href@noop {}
  {\enquote {\bibinfo {title} {Material decomposition from a single x-ray
  projection via single-grid phase contrast imaging},}\ } (\bibinfo {year}
  {2020}),\ \Eprint {http://arxiv.org/abs/2002.04417} {arXiv:2002.04417}
  \BibitemShut {NoStop}%
\bibitem [{\citenamefont {{Sanchez del Rio}}\ and\ \citenamefont
  {Dejus}(2011)}]{delRio2011}%
  \BibitemOpen
  \bibfield  {author} {\bibinfo {author} {\bibfnamefont {M.}~\bibnamefont
  {{Sanchez del Rio}}}\ and\ \bibinfo {author} {\bibfnamefont {R.~J.}\
  \bibnamefont {Dejus}},\ }\bibfield  {title} {\enquote {\bibinfo {title}
  {{XOP} v2.4: recent developments of the x-ray optics software toolkit},}\
  }\href@noop {} {\bibfield  {journal} {\bibinfo  {journal} {Proc. SPIE}\
  }\textbf {\bibinfo {volume} {8141}},\ \bibinfo {pages} {814115} (\bibinfo
  {year} {2011})}\BibitemShut {NoStop}%
\bibitem [{\citenamefont {Gureyev}\ \emph {et~al.}(2002)\citenamefont
  {Gureyev}, \citenamefont {Stevenson}, \citenamefont {Paganin}, \citenamefont
  {Weitkamp}, \citenamefont {Snigirev}, \citenamefont {Snigireva},\ and\
  \citenamefont {Wilkins}}]{Gureyev2002}%
  \BibitemOpen
  \bibfield  {author} {\bibinfo {author} {\bibfnamefont {T.~E.}\ \bibnamefont
  {Gureyev}}, \bibinfo {author} {\bibfnamefont {A.~W.}\ \bibnamefont
  {Stevenson}}, \bibinfo {author} {\bibfnamefont {D.~M.}\ \bibnamefont
  {Paganin}}, \bibinfo {author} {\bibfnamefont {T.}~\bibnamefont {Weitkamp}},
  \bibinfo {author} {\bibfnamefont {A.}~\bibnamefont {Snigirev}}, \bibinfo
  {author} {\bibfnamefont {I.}~\bibnamefont {Snigireva}}, \ and\ \bibinfo
  {author} {\bibfnamefont {S.~W.}\ \bibnamefont {Wilkins}},\ }\bibfield
  {title} {\enquote {\bibinfo {title} {Quantitative analysis of two-component
  samples using in-line hard x-ray images},}\ }\href@noop {} {\bibfield
  {journal} {\bibinfo  {journal} {J. Synchrotron Radiat.}\ }\textbf {\bibinfo
  {volume} {9}},\ \bibinfo {pages} {148--153} (\bibinfo {year}
  {2002})}\BibitemShut {NoStop}%
\bibitem [{\citenamefont {Subbarao}\ \emph {et~al.}(1995)\citenamefont
  {Subbarao}, \citenamefont {Wei},\ and\ \citenamefont {Surya}}]{Unsharp0}%
  \BibitemOpen
  \bibfield  {author} {\bibinfo {author} {\bibfnamefont {M.}~\bibnamefont
  {Subbarao}}, \bibinfo {author} {\bibfnamefont {T.-C.}\ \bibnamefont {Wei}}, \
  and\ \bibinfo {author} {\bibfnamefont {G.}~\bibnamefont {Surya}},\ }\bibfield
   {title} {\enquote {\bibinfo {title} {Focused image recovery from two
  defocused images recorded with different camera settings},}\ }\href@noop {}
  {\bibfield  {journal} {\bibinfo  {journal} {IEEE Trans. Image Process.}\
  }\textbf {\bibinfo {volume} {4}},\ \bibinfo {pages} {1613--1628} (\bibinfo
  {year} {1995})}\BibitemShut {NoStop}%
\bibitem [{\citenamefont {Gureyev}\ \emph {et~al.}(2004)\citenamefont
  {Gureyev}, \citenamefont {Stevenson}, \citenamefont {Nesterets},\ and\
  \citenamefont {Wilkins}}]{DeblurByDefocus}%
  \BibitemOpen
  \bibfield  {author} {\bibinfo {author} {\bibfnamefont {T.~E.}\ \bibnamefont
  {Gureyev}}, \bibinfo {author} {\bibfnamefont {A.~W.}\ \bibnamefont
  {Stevenson}}, \bibinfo {author} {\bibfnamefont {{\relax Ya}.~I.}\
  \bibnamefont {Nesterets}}, \ and\ \bibinfo {author} {\bibfnamefont {S.~W.}\
  \bibnamefont {Wilkins}},\ }\bibfield  {title} {\enquote {\bibinfo {title}
  {Image deblurring by means of defocus},}\ }\href@noop {} {\bibfield
  {journal} {\bibinfo  {journal} {Opt. Commun.}\ }\textbf {\bibinfo {volume}
  {240}},\ \bibinfo {pages} {81--88} (\bibinfo {year} {2004})}\BibitemShut
  {NoStop}%
\bibitem [{\citenamefont {Beltran}\ \emph {et~al.}(2018)\citenamefont
  {Beltran}, \citenamefont {Paganin},\ and\ \citenamefont
  {Pelliccia}}]{beltran2018}%
  \BibitemOpen
  \bibfield  {author} {\bibinfo {author} {\bibfnamefont {M.~A.}\ \bibnamefont
  {Beltran}}, \bibinfo {author} {\bibfnamefont {D.~M.}\ \bibnamefont
  {Paganin}}, \ and\ \bibinfo {author} {\bibfnamefont {D.}~\bibnamefont
  {Pelliccia}},\ }\bibfield  {title} {\enquote {\bibinfo {title}
  {Phase-and-amplitude recovery from a single phase-contrast image using
  partially spatially coherent x-ray radiation},}\ }\href@noop {} {\bibfield
  {journal} {\bibinfo  {journal} {J. Opt.}\ }\textbf {\bibinfo {volume} {20}},\
  \bibinfo {pages} {055605} (\bibinfo {year} {2018})}\BibitemShut {NoStop}%
\bibitem [{\citenamefont {Mittone}\ \emph {et~al.}(2017)\citenamefont
  {Mittone}, \citenamefont {Manakov}, \citenamefont {Broche}, \citenamefont
  {Jarnias}, \citenamefont {Coan},\ and\ \citenamefont {Bravin}}]{Mittone2017}%
  \BibitemOpen
  \bibfield  {author} {\bibinfo {author} {\bibfnamefont {A.}~\bibnamefont
  {Mittone}}, \bibinfo {author} {\bibfnamefont {I.}~\bibnamefont {Manakov}},
  \bibinfo {author} {\bibfnamefont {L.}~\bibnamefont {Broche}}, \bibinfo
  {author} {\bibfnamefont {C.}~\bibnamefont {Jarnias}}, \bibinfo {author}
  {\bibfnamefont {P.}~\bibnamefont {Coan}}, \ and\ \bibinfo {author}
  {\bibfnamefont {A.}~\bibnamefont {Bravin}},\ }\bibfield  {title} {\enquote
  {\bibinfo {title} {{Characterization of a sCMOS-based high-resolution imaging
  system}},}\ }\href {\doibase 10.1107/S160057751701222X} {\bibfield  {journal}
  {\bibinfo  {journal} {J. Synchrotron Radiat.}\ }\textbf {\bibinfo {volume}
  {24}},\ \bibinfo {pages} {1226--1236} (\bibinfo {year} {2017})}\BibitemShut
  {NoStop}%
\bibitem [{\citenamefont {Risken}(1989)}]{Risken1989}%
  \BibitemOpen
  \bibfield  {author} {\bibinfo {author} {\bibfnamefont {H.}~\bibnamefont
  {Risken}},\ }\href@noop {} {\emph {\bibinfo {title} {The Fokker--Planck
  Equation: Methods of Solution and Applications}}},\ \bibinfo {edition} {2nd}\
  ed.\ (\bibinfo  {publisher} {Springer Verlag, Berlin},\ \bibinfo {year}
  {1989})\BibitemShut {NoStop}%
\bibitem [{\citenamefont {Morgan}\ and\ \citenamefont
  {Paganin}(2019)}]{MorganPaganin2019}%
  \BibitemOpen
  \bibfield  {author} {\bibinfo {author} {\bibfnamefont {K.~S.}\ \bibnamefont
  {Morgan}}\ and\ \bibinfo {author} {\bibfnamefont {D.~M.}\ \bibnamefont
  {Paganin}},\ }\bibfield  {title} {\enquote {\bibinfo {title} {Applying the
  {F}okker--{P}lanck equation to x-ray grating-based phase and dark-field
  imaging},}\ }\href@noop {} {\bibfield  {journal} {\bibinfo  {journal} {Sci.
  Rep.}\ }\textbf {\bibinfo {volume} {9}},\ \bibinfo {pages} {17465} (\bibinfo
  {year} {2019})}\BibitemShut {NoStop}%
\bibitem [{\citenamefont {Paganin}\ and\ \citenamefont
  {Morgan}(2019)}]{PaganinMorgan2019}%
  \BibitemOpen
  \bibfield  {author} {\bibinfo {author} {\bibfnamefont {D.~M.}\ \bibnamefont
  {Paganin}}\ and\ \bibinfo {author} {\bibfnamefont {K.~S.}\ \bibnamefont
  {Morgan}},\ }\bibfield  {title} {\enquote {\bibinfo {title} {X-ray
  {F}okker--{P}lanck equation for paraxial imaging},}\ }\href@noop {}
  {\bibfield  {journal} {\bibinfo  {journal} {Sci. Rep.}\ }\textbf {\bibinfo
  {volume} {9}},\ \bibinfo {pages} {17537} (\bibinfo {year}
  {2019})}\BibitemShut {NoStop}%
\bibitem [{\citenamefont {Berujon}\ and\ \citenamefont
  {Ziegler}(2016)}]{Berujon2015c}%
  \BibitemOpen
  \bibfield  {author} {\bibinfo {author} {\bibfnamefont {S.}~\bibnamefont
  {Berujon}}\ and\ \bibinfo {author} {\bibfnamefont {E.}~\bibnamefont
  {Ziegler}},\ }\bibfield  {title} {\enquote {\bibinfo {title} {X-ray
  multimodal tomography using speckle-vector tracking},}\ }\href@noop {}
  {\bibfield  {journal} {\bibinfo  {journal} {Phys. Rev. Appl.}\ }\textbf
  {\bibinfo {volume} {5}},\ \bibinfo {pages} {044014} (\bibinfo {year}
  {2016})}\BibitemShut {NoStop}%
\bibitem [{\citenamefont {Zdora}\ \emph {et~al.}(2017)\citenamefont {Zdora},
  \citenamefont {Thibault}, \citenamefont {Zhou}, \citenamefont {Koch},
  \citenamefont {Romell}, \citenamefont {Sala}, \citenamefont {Last},
  \citenamefont {Rau},\ and\ \citenamefont {Zanette}}]{Zdora2017}%
  \BibitemOpen
  \bibfield  {author} {\bibinfo {author} {\bibfnamefont {M.-C.}\ \bibnamefont
  {Zdora}}, \bibinfo {author} {\bibfnamefont {P.}~\bibnamefont {Thibault}},
  \bibinfo {author} {\bibfnamefont {T.}~\bibnamefont {Zhou}}, \bibinfo {author}
  {\bibfnamefont {F.~J.}\ \bibnamefont {Koch}}, \bibinfo {author}
  {\bibfnamefont {J.}~\bibnamefont {Romell}}, \bibinfo {author} {\bibfnamefont
  {S.}~\bibnamefont {Sala}}, \bibinfo {author} {\bibfnamefont {A.}~\bibnamefont
  {Last}}, \bibinfo {author} {\bibfnamefont {C.}~\bibnamefont {Rau}}, \ and\
  \bibinfo {author} {\bibfnamefont {I.}~\bibnamefont {Zanette}},\ }\bibfield
  {title} {\enquote {\bibinfo {title} {{X-ray Phase-Contrast Imaging and
  Metrology through Unified Modulated Pattern Analysis}},}\ }\href {\doibase
  10.1103/PhysRevLett.118.203903} {\bibfield  {journal} {\bibinfo  {journal}
  {Phys. Rev. Lett.}\ }\textbf {\bibinfo {volume} {118}},\ \bibinfo {pages}
  {203903} (\bibinfo {year} {2017})}\BibitemShut {NoStop}%
\bibitem [{\citenamefont {Stepanov}()}]{StepanovWebsite}%
  \BibitemOpen
  \bibfield  {author} {\bibinfo {author} {\bibfnamefont {Sergey}\ \bibnamefont
  {Stepanov}},\ }\href@noop {} {\enquote {\bibinfo {title} {X-ray server},}\
  }\bibinfo {howpublished} {\url{https://x-server.gmca.aps.anl.gov/}},\
  \bibinfo {note} {accessed November 8, 2019}\BibitemShut {NoStop}%
\bibitem [{\citenamefont {Mirone}\ \emph {et~al.}(2014)\citenamefont {Mirone},
  \citenamefont {Brun}, \citenamefont {Gouillart}, \citenamefont {Tafforeau},\
  and\ \citenamefont {Kieffer}}]{mirone2014}%
  \BibitemOpen
  \bibfield  {author} {\bibinfo {author} {\bibfnamefont {A.}~\bibnamefont
  {Mirone}}, \bibinfo {author} {\bibfnamefont {E.}~\bibnamefont {Brun}},
  \bibinfo {author} {\bibfnamefont {E.}~\bibnamefont {Gouillart}}, \bibinfo
  {author} {\bibfnamefont {P.}~\bibnamefont {Tafforeau}}, \ and\ \bibinfo
  {author} {\bibfnamefont {J.}~\bibnamefont {Kieffer}},\ }\bibfield  {title}
  {\enquote {\bibinfo {title} {The {PyHST2} hybrid distributed code for high
  speed tomographic reconstruction with iterative reconstruction and a priori
  knowledge capabilities},}\ }\href@noop {} {\bibfield  {journal} {\bibinfo
  {journal} {Nucl. Instrum. Meth. Phys. Res. B}\ }\textbf {\bibinfo {volume}
  {324}},\ \bibinfo {pages} {41--48} (\bibinfo {year} {2014})}\BibitemShut
  {NoStop}%
\bibitem [{\citenamefont {Mayo}\ and\ \citenamefont
  {Sexton}(2004)}]{MayoSexton2004}%
  \BibitemOpen
  \bibfield  {author} {\bibinfo {author} {\bibfnamefont {S.~C.}\ \bibnamefont
  {Mayo}}\ and\ \bibinfo {author} {\bibfnamefont {B.}~\bibnamefont {Sexton}},\
  }\bibfield  {title} {\enquote {\bibinfo {title} {Refractive microlens array
  for wave-front analysis in the medium to hard x-ray range},}\ }\href@noop {}
  {\bibfield  {journal} {\bibinfo  {journal} {Opt. Lett.}\ }\textbf {\bibinfo
  {volume} {29}},\ \bibinfo {pages} {866--868} (\bibinfo {year}
  {2004})}\BibitemShut {NoStop}%
\bibitem [{\citenamefont {Berto}\ \emph {et~al.}(2017)\citenamefont {Berto},
  \citenamefont {Rigneault},\ and\ \citenamefont {Guillon}}]{Berto2017}%
  \BibitemOpen
  \bibfield  {author} {\bibinfo {author} {\bibfnamefont {P.}~\bibnamefont
  {Berto}}, \bibinfo {author} {\bibfnamefont {H.}~\bibnamefont {Rigneault}}, \
  and\ \bibinfo {author} {\bibfnamefont {M.}~\bibnamefont {Guillon}},\
  }\bibfield  {title} {\enquote {\bibinfo {title} {Wavefront sensing with a
  thin diffuser},}\ }\href {\doibase 10.1364/OL.42.005117} {\bibfield
  {journal} {\bibinfo  {journal} {Opt. Lett.}\ }\textbf {\bibinfo {volume}
  {42}},\ \bibinfo {pages} {5117--5120} (\bibinfo {year} {2017})}\BibitemShut
  {NoStop}%
\bibitem [{\citenamefont {Massig}(1999)}]{Massig1}%
  \BibitemOpen
  \bibfield  {author} {\bibinfo {author} {\bibfnamefont {J.~H.}\ \bibnamefont
  {Massig}},\ }\bibfield  {title} {\enquote {\bibinfo {title} {Measurement of
  phase objects by simple means},}\ }\href@noop {} {\bibfield  {journal}
  {\bibinfo  {journal} {Appl. Opt.}\ }\textbf {\bibinfo {volume} {38}},\
  \bibinfo {pages} {4103--4105} (\bibinfo {year} {1999})}\BibitemShut {NoStop}%
\bibitem [{\citenamefont {Massig}(2001)}]{Massig2}%
  \BibitemOpen
  \bibfield  {author} {\bibinfo {author} {\bibfnamefont {J.~H.}\ \bibnamefont
  {Massig}},\ }\bibfield  {title} {\enquote {\bibinfo {title} {Deformation
  measurement on specular surfaces by simple means},}\ }\href@noop {}
  {\bibfield  {journal} {\bibinfo  {journal} {Opt. Eng.}\ }\textbf {\bibinfo
  {volume} {40}},\ \bibinfo {pages} {2315--2318} (\bibinfo {year}
  {2001})}\BibitemShut {NoStop}%
\bibitem [{\citenamefont {Perciante}\ and\ \citenamefont
  {Ferrari}(2000)}]{Perciante}%
  \BibitemOpen
  \bibfield  {author} {\bibinfo {author} {\bibfnamefont {C.~D.}\ \bibnamefont
  {Perciante}}\ and\ \bibinfo {author} {\bibfnamefont {J.~A.}\ \bibnamefont
  {Ferrari}},\ }\bibfield  {title} {\enquote {\bibinfo {title} {Visualization
  of two-dimensional phase gradients by subtraction of a reference periodic
  pattern},}\ }\href@noop {} {\bibfield  {journal} {\bibinfo  {journal} {Appl.
  Opt.}\ }\textbf {\bibinfo {volume} {39}},\ \bibinfo {pages} {2081--2083}
  (\bibinfo {year} {2000})}\BibitemShut {NoStop}%
\bibitem [{\citenamefont {Clark}\ \emph {et~al.}(2019)\citenamefont {Clark},
  \citenamefont {Petersen}, \citenamefont {Williams}, \citenamefont {Morgan},
  \citenamefont {Paganin},\ and\ \citenamefont {Findlay}}]{Clark2019}%
  \BibitemOpen
  \bibfield  {author} {\bibinfo {author} {\bibfnamefont {L.}~\bibnamefont
  {Clark}}, \bibinfo {author} {\bibfnamefont {T.~C.}\ \bibnamefont {Petersen}},
  \bibinfo {author} {\bibfnamefont {T.}~\bibnamefont {Williams}}, \bibinfo
  {author} {\bibfnamefont {M.~J.}\ \bibnamefont {Morgan}}, \bibinfo {author}
  {\bibfnamefont {D.~M.}\ \bibnamefont {Paganin}}, \ and\ \bibinfo {author}
  {\bibfnamefont {S.~D.}\ \bibnamefont {Findlay}},\ }\bibfield  {title}
  {\enquote {\bibinfo {title} {High contrast at low dose using a single,
  defocussed transmission electron micrograph},}\ }\href@noop {} {\bibfield
  {journal} {\bibinfo  {journal} {Micron}\ }\textbf {\bibinfo {volume} {124}},\
  \bibinfo {pages} {102701} (\bibinfo {year} {2019})}\BibitemShut {NoStop}%
\bibitem [{\citenamefont {Beltran}\ \emph {et~al.}(2010)\citenamefont
  {Beltran}, \citenamefont {Paganin}, \citenamefont {Uesugi},\ and\
  \citenamefont {Kitchen}}]{beltran2010}%
  \BibitemOpen
  \bibfield  {author} {\bibinfo {author} {\bibfnamefont {M.~A.}\ \bibnamefont
  {Beltran}}, \bibinfo {author} {\bibfnamefont {D.~M.}\ \bibnamefont
  {Paganin}}, \bibinfo {author} {\bibfnamefont {K.}~\bibnamefont {Uesugi}}, \
  and\ \bibinfo {author} {\bibfnamefont {M.~J.}\ \bibnamefont {Kitchen}},\
  }\bibfield  {title} {\enquote {\bibinfo {title} {{2D} and {3D} {X}-ray phase
  retrieval of multi-material objects using a single defocus distance},}\
  }\href {http://www.opticsexpress.org/abstract.cfm?URI=oe-18-7-6423}
  {\bibfield  {journal} {\bibinfo  {journal} {Opt. Express}\ }\textbf {\bibinfo
  {volume} {18}},\ \bibinfo {pages} {6423--6436} (\bibinfo {year}
  {2010})}\BibitemShut {NoStop}%
\bibitem [{\citenamefont {Beltran}\ \emph {et~al.}(2011)\citenamefont
  {Beltran}, \citenamefont {Paganin}, \citenamefont {Siu}, \citenamefont
  {Fouras}, \citenamefont {Hooper}, \citenamefont {Reser},\ and\ \citenamefont
  {Kitchen}}]{beltran2011}%
  \BibitemOpen
  \bibfield  {author} {\bibinfo {author} {\bibfnamefont {M.~A.}\ \bibnamefont
  {Beltran}}, \bibinfo {author} {\bibfnamefont {D.~M.}\ \bibnamefont
  {Paganin}}, \bibinfo {author} {\bibfnamefont {K.~K.~W.}\ \bibnamefont {Siu}},
  \bibinfo {author} {\bibfnamefont {A.}~\bibnamefont {Fouras}}, \bibinfo
  {author} {\bibfnamefont {S.~B.}\ \bibnamefont {Hooper}}, \bibinfo {author}
  {\bibfnamefont {D.~H.}\ \bibnamefont {Reser}}, \ and\ \bibinfo {author}
  {\bibfnamefont {M.~J.}\ \bibnamefont {Kitchen}},\ }\bibfield  {title}
  {\enquote {\bibinfo {title} {Interface-specific x-ray phase retrieval
  tomography of complex biological organs},}\ }\href@noop {} {\bibfield
  {journal} {\bibinfo  {journal} {Phys. Med. Biol.}\ }\textbf {\bibinfo
  {volume} {56}},\ \bibinfo {pages} {7353--7369} (\bibinfo {year}
  {2011})}\BibitemShut {NoStop}%
\bibitem [{\citenamefont {Nesterets}\ and\ \citenamefont
  {Gureyev}(2014)}]{SNRboost3}%
  \BibitemOpen
  \bibfield  {author} {\bibinfo {author} {\bibfnamefont {Ya.~I.}\ \bibnamefont
  {Nesterets}}\ and\ \bibinfo {author} {\bibfnamefont {T.~E.}\ \bibnamefont
  {Gureyev}},\ }\bibfield  {title} {\enquote {\bibinfo {title} {Noise
  propagation in x-ray phase-contrast imaging and computed tomography},}\
  }\href@noop {} {\bibfield  {journal} {\bibinfo  {journal} {J. Phys. D: Appl.
  Phys.}\ }\textbf {\bibinfo {volume} {47}},\ \bibinfo {pages} {105402}
  (\bibinfo {year} {2014})}\BibitemShut {NoStop}%
\bibitem [{\citenamefont {Gureyev}\ \emph {et~al.}(2014)\citenamefont
  {Gureyev}, \citenamefont {Mayo}, \citenamefont {Nesterets}, \citenamefont
  {Mohammadi}, \citenamefont {Lockie}, \citenamefont {Menk}, \citenamefont
  {Arfelli}, \citenamefont {Pavlov}, \citenamefont {Kitchen}, \citenamefont
  {Zanconati}, \citenamefont {Dullin},\ and\ \citenamefont
  {Tromba}}]{SNRboost4}%
  \BibitemOpen
  \bibfield  {author} {\bibinfo {author} {\bibfnamefont {T.~E.}\ \bibnamefont
  {Gureyev}}, \bibinfo {author} {\bibfnamefont {S.~C.}\ \bibnamefont {Mayo}},
  \bibinfo {author} {\bibfnamefont {Ya.~I.}\ \bibnamefont {Nesterets}},
  \bibinfo {author} {\bibfnamefont {S.}~\bibnamefont {Mohammadi}}, \bibinfo
  {author} {\bibfnamefont {D.}~\bibnamefont {Lockie}}, \bibinfo {author}
  {\bibfnamefont {R.~H.}\ \bibnamefont {Menk}}, \bibinfo {author}
  {\bibfnamefont {F.}~\bibnamefont {Arfelli}}, \bibinfo {author} {\bibfnamefont
  {K.~M.}\ \bibnamefont {Pavlov}}, \bibinfo {author} {\bibfnamefont {M.~J.}\
  \bibnamefont {Kitchen}}, \bibinfo {author} {\bibfnamefont {F.}~\bibnamefont
  {Zanconati}}, \bibinfo {author} {\bibfnamefont {C.}~\bibnamefont {Dullin}}, \
  and\ \bibinfo {author} {\bibfnamefont {G.}~\bibnamefont {Tromba}},\
  }\bibfield  {title} {\enquote {\bibinfo {title} {Investigation of the imaging
  quality of synchrotron-based phase-contrast mammographic tomography},}\
  }\href@noop {} {\bibfield  {journal} {\bibinfo  {journal} {J. Phys. D: Appl.
  Phys.}\ }\textbf {\bibinfo {volume} {47}},\ \bibinfo {pages} {365401}
  (\bibinfo {year} {2014})}\BibitemShut {NoStop}%
\bibitem [{\citenamefont {Kitchen}\ \emph {et~al.}(2017)\citenamefont
  {Kitchen}, \citenamefont {Buckley}, \citenamefont {Gureyev}, \citenamefont
  {Wallace}, \citenamefont {Andres-{T}hio}, \citenamefont {Uesugi},
  \citenamefont {Yagi},\ and\ \citenamefont {Hooper}}]{SNRboost5}%
  \BibitemOpen
  \bibfield  {author} {\bibinfo {author} {\bibfnamefont {M.~J.}\ \bibnamefont
  {Kitchen}}, \bibinfo {author} {\bibfnamefont {G.~A.}\ \bibnamefont
  {Buckley}}, \bibinfo {author} {\bibfnamefont {T.~E.}\ \bibnamefont
  {Gureyev}}, \bibinfo {author} {\bibfnamefont {M.~J.}\ \bibnamefont
  {Wallace}}, \bibinfo {author} {\bibfnamefont {N.}~\bibnamefont
  {Andres-{T}hio}}, \bibinfo {author} {\bibfnamefont {K.}~\bibnamefont
  {Uesugi}}, \bibinfo {author} {\bibfnamefont {N.}~\bibnamefont {Yagi}}, \ and\
  \bibinfo {author} {\bibfnamefont {S.~B.}\ \bibnamefont {Hooper}},\ }\bibfield
   {title} {\enquote {\bibinfo {title} {{CT} dose reduction factors in the
  thousands using x-ray phase contrast},}\ }\href@noop {} {\bibfield  {journal}
  {\bibinfo  {journal} {Sci. Rep.}\ }\textbf {\bibinfo {volume} {7}},\ \bibinfo
  {pages} {15953} (\bibinfo {year} {2017})}\BibitemShut {NoStop}%
\bibitem [{\citenamefont {Gureyev}\ \emph {et~al.}(2017)\citenamefont
  {Gureyev}, \citenamefont {Nesterets}, \citenamefont {Kozlov}, \citenamefont
  {Paganin},\ and\ \citenamefont {Quiney}}]{gureyev2017unreasonable}%
  \BibitemOpen
  \bibfield  {author} {\bibinfo {author} {\bibfnamefont {T.~E.}\ \bibnamefont
  {Gureyev}}, \bibinfo {author} {\bibfnamefont {{Ya}.~I.}\ \bibnamefont
  {Nesterets}}, \bibinfo {author} {\bibfnamefont {A.}~\bibnamefont {Kozlov}},
  \bibinfo {author} {\bibfnamefont {D.~M.}\ \bibnamefont {Paganin}}, \ and\
  \bibinfo {author} {\bibfnamefont {H.~M.}\ \bibnamefont {Quiney}},\ }\bibfield
   {title} {\enquote {\bibinfo {title} {On the `unreasonable' effectiveness of
  {T}ransport of {I}ntensity imaging and optical deconvolution},}\ }\href@noop
  {} {\bibfield  {journal} {\bibinfo  {journal} {J. Opt. Soc. Am. A}\ }\textbf
  {\bibinfo {volume} {34}},\ \bibinfo {pages} {2251--2260} (\bibinfo {year}
  {2017})}\BibitemShut {NoStop}%
\bibitem [{\citenamefont {Kemp}(2018)}]{kemp2018propagation}%
  \BibitemOpen
  \bibfield  {author} {\bibinfo {author} {\bibfnamefont {Z.~D.~C.}\
  \bibnamefont {Kemp}},\ }\bibfield  {title} {\enquote {\bibinfo {title}
  {Propagation based phase retrieval of simulated intensity measurements using
  artificial neural networks},}\ }\href@noop {} {\bibfield  {journal} {\bibinfo
   {journal} {J. Opt.}\ }\textbf {\bibinfo {volume} {20}},\ \bibinfo {pages}
  {045606} (\bibinfo {year} {2018})}\BibitemShut {NoStop}%
\bibitem [{\citenamefont {Rivenson}\ \emph {et~al.}(2018)\citenamefont
  {Rivenson}, \citenamefont {Zhang}, \citenamefont {G{\"u}nayd{\i}n},
  \citenamefont {Teng},\ and\ \citenamefont {Ozcan}}]{rivenson2018phase}%
  \BibitemOpen
  \bibfield  {author} {\bibinfo {author} {\bibfnamefont {Y.}~\bibnamefont
  {Rivenson}}, \bibinfo {author} {\bibfnamefont {Y.}~\bibnamefont {Zhang}},
  \bibinfo {author} {\bibfnamefont {H.}~\bibnamefont {G{\"u}nayd{\i}n}},
  \bibinfo {author} {\bibfnamefont {D.}~\bibnamefont {Teng}}, \ and\ \bibinfo
  {author} {\bibfnamefont {A.}~\bibnamefont {Ozcan}},\ }\bibfield  {title}
  {\enquote {\bibinfo {title} {Phase recovery and holographic image
  reconstruction using deep learning in neural networks},}\ }\href@noop {}
  {\bibfield  {journal} {\bibinfo  {journal} {Light Sci. Appl.}\ }\textbf
  {\bibinfo {volume} {7}},\ \bibinfo {pages} {17141} (\bibinfo {year}
  {2018})}\BibitemShut {NoStop}%
\bibitem [{\citenamefont {Jiang}\ \emph {et~al.}(2018)\citenamefont {Jiang},
  \citenamefont {Guo}, \citenamefont {Liao},\ and\ \citenamefont
  {Zheng}}]{jiang2018solving}%
  \BibitemOpen
  \bibfield  {author} {\bibinfo {author} {\bibfnamefont {S.}~\bibnamefont
  {Jiang}}, \bibinfo {author} {\bibfnamefont {K.}~\bibnamefont {Guo}}, \bibinfo
  {author} {\bibfnamefont {J.}~\bibnamefont {Liao}}, \ and\ \bibinfo {author}
  {\bibfnamefont {G.}~\bibnamefont {Zheng}},\ }\bibfield  {title} {\enquote
  {\bibinfo {title} {Solving {F}ourier ptychographic imaging problems via
  neural network modeling and {T}ensor{F}low},}\ }\href@noop {} {\bibfield
  {journal} {\bibinfo  {journal} {Biomed. Opt. Express}\ }\textbf {\bibinfo
  {volume} {9}},\ \bibinfo {pages} {3306--3319} (\bibinfo {year}
  {2018})}\BibitemShut {NoStop}%
\bibitem [{\citenamefont {Li}\ \emph {et~al.}(2018)\citenamefont {Li},
  \citenamefont {Kingston}, \citenamefont {Myers}, \citenamefont {Beeching},\
  and\ \citenamefont {Sheppard}}]{Li:18}%
  \BibitemOpen
  \bibfield  {author} {\bibinfo {author} {\bibfnamefont {H.}~\bibnamefont
  {Li}}, \bibinfo {author} {\bibfnamefont {A.~M.}\ \bibnamefont {Kingston}},
  \bibinfo {author} {\bibfnamefont {G.~R.}\ \bibnamefont {Myers}}, \bibinfo
  {author} {\bibfnamefont {L.}~\bibnamefont {Beeching}}, \ and\ \bibinfo
  {author} {\bibfnamefont {A.~P.}\ \bibnamefont {Sheppard}},\ }\bibfield
  {title} {\enquote {\bibinfo {title} {Linear iterative near-field phase
  retrieval ({LIPR}) for dual-energy x-ray imaging and material
  discrimination},}\ }\href@noop {} {\bibfield  {journal} {\bibinfo  {journal}
  {J. Opt. Soc. Am. A}\ }\textbf {\bibinfo {volume} {35}},\ \bibinfo {pages}
  {A30--A39} (\bibinfo {year} {2018})}\BibitemShut {NoStop}%
\bibitem [{\citenamefont {G\"{u}rsoy}\ and\ \citenamefont
  {Das}(2013)}]{Gursoy:13}%
  \BibitemOpen
  \bibfield  {author} {\bibinfo {author} {\bibfnamefont {D.}~\bibnamefont
  {G\"{u}rsoy}}\ and\ \bibinfo {author} {\bibfnamefont {M.}~\bibnamefont
  {Das}},\ }\bibfield  {title} {\enquote {\bibinfo {title} {Single-step
  absorption and phase retrieval with polychromatic x rays using a spectral
  detector},}\ }\href@noop {} {\bibfield  {journal} {\bibinfo  {journal} {Opt.
  Lett.}\ }\textbf {\bibinfo {volume} {38}},\ \bibinfo {pages} {1461--1463}
  (\bibinfo {year} {2013})}\BibitemShut {NoStop}%
\end{thebibliography}%

\end{document}